\documentclass[11pt]{article}

\usepackage{booktabs}
\usepackage[margin=15pt,font=small,labelfont={bf,sf},justification=justified]{caption}
\usepackage{color}
\usepackage{float}
\usepackage{graphicx}
\usepackage{lscape}
\usepackage{mathtools}
\usepackage{multirow}
\usepackage{tabularx}
\usepackage{titleref}
\usepackage{url}
\usepackage{hyperref} 
\usepackage[numbers,sort&compress]{natbib}
\usepackage{multibib}
\newcites{S}{Supplementary References}

\newif\ifbembo
\newif\ifcharter
\newif\iferewhon
\newif\iflibertine
\newif\iflibertinealt
\newif\ifpalantino
\newif\iftimesnewroman

\bembofalse
\chartertrue
\erewhonfalse
\libertinefalse
\libertinealtfalse
\palantinofalse
\timesnewromanfalse

\ifbembo
  \usepackage[p,osf]{ETbb} 
  \usepackage[scaled=.95,type1]{cabin} 
  \usepackage[varqu,varl]{zi4}
  \usepackage[T1]{fontenc}
  \usepackage[libertine,vvarbb]{newtxmath}
  \usepackage[cal=boondoxo,bb=boondox]{mathalfa}
\fi

\ifcharter
  \usepackage[scaled=.98,p]{XCharter}
  \usepackage[scaled=1.04,varqu,varl]{inconsolata}
  \usepackage[type1]{cabin}
  \usepackage[uprightscript,xcharter,vvarbb,scaled=1.05]{newtxmath}
  \usepackage[cal=euler,scr=boondoxo,bb=boondox]{mathalpha}  
\fi

\iferewhon
  \usepackage[p,osf,scaled=.98,space]{erewhon} 
  \usepackage[varqu,varl]{inconsolata} 
  \usepackage[type1,scaled=.95]{cabin} 
  \usepackage[utopia,vvarbb]{newtxmath}
\fi

\iflibertine
  \usepackage[sb]{libertinus}
  \usepackage[T1]{fontenc}
  \usepackage{textcomp}
  \usepackage[varqu,varl]{zi4}
  \usepackage{libertinust1math} 
  \usepackage[scr=boondoxo,bb=boondox]{mathalpha} 
\fi

\iflibertinealt
  \usepackage[sb]{libertinus}
  \usepackage[T1]{fontenc}
  \usepackage{textcomp}
  \usepackage{libertinust1math} 
  \usepackage[scr=boondoxo,bb=boondox]{mathalpha} 
\fi

\ifpalantino
  \usepackage[largesc]{newpxtext}
  \usepackage[vvarbb]{newpxmath}
\fi

\usepackage[T1]{fontenc}
\usepackage[final,protrusion=true,expansion=true]{microtype}

\usepackage{titling}
\usepackage{authblk}
\pretitle{\begin{center}\bfseries\sffamily\LARGE}
\posttitle{\end{center}}
\usepackage{abstract}

\usepackage{sectsty} 
\allsectionsfont{\sffamily}

\usepackage[left=1in,right=1in,top=1in,bottom=1in,includefoot,heightrounded]{geometry}

\makeatletter
\patchcmd{\LS@rot}{90}{-90}{}{}
\patchcmd{\endlandscape}{90}{-90}{}{}
\makeatother

\newcommand{\beginsupplement}{%
        \setcounter{table}{0}
        \renewcommand{\thetable}{S\arabic{table}}%
        \setcounter{figure}{0}
        \renewcommand{\thefigure}{S\arabic{figure}}%
        \setcounter{section}{0}
        \renewcommand{\thesection}{S\arabic{section}}
        \setcounter{equation}{0}
        \renewcommand{\theequation}{S\arabic{equation}}     
}

\title{Sensitivity of ECG QRS Complexes to His-Purkinje Structure in Computational Heart Models}
\author[a,b,*]{Preetam V. Tanikella}
\author[a]{Laryssa Abdala}
\author[a]{Karin Leiderman}
\author[b]{Annie Green Howard}
\author[a, c, d, e]{Boyce E. Griffith}
\affil[a]{Department of Mathematics, University North Carolina, Chapel Hill, NC 27599, USA}
\affil[b]{Department of Biostatistics,  University North Carolina, Chapel Hill, NC 27599, USA}
\affil[c]{Department of Biomedical Engineering, University of North Carolina, Chapel Hill, NC 27599, USA}
\affil[d]{Carolina Center for Interdisciplinary Applied Mathematics, University of North Carolina, Chapel Hill, NC 27599, USA }
\affil[e]{Computational Medicine Program, University of North Carolina School of Medicine, Chapel Hill, NC 27599, USA}
\affil[*]{To whom correspondence should be addressed: Email: ptanikella@unc.edu}

\begin{document}
\maketitle 
\begin{abstract}
\noindent Cardiac digital twins (CDT) are emerging as a potentially transformative tool in cardiology. A critical yet understudied determinant of CDT accuracy is the His-Purkinje system (HPS), which influences ventricular depolarization and shapes the QRS complex of the electrocardiogram (ECG). Here, we quantify how structural variations in the HPS alter QRS morphology and identify which parameters drive this variability. We generated HPS structures using a fractal-tree, rule-based algorithm, systematically varying nine model parameters and assessing their effects on ten QRS-related metrics. We conducted a Sobol sensitivity analysis to quantify direct and interaction-driven contributions of each parameter to observed variability. Our results suggest that most minor changes in HPS structure exert minimal influence on individual QRS features; however, certain parameter combinations can produce abnormal QRS morphologies. Wave durations and peak amplitudes of the QRS complex exhibit low sensitivity to individual HPS parameter variations; however, we found that specific parameter combinations can result in interactions that significantly alter these aspects of QRS morphology. We found that certain HPS structures can cause premature QRS formation, obscuring P-wave formation. QRS timing variability was primarily driven by interactions among branch and fascicle angles and branch repulsivity, though other parameters also showed notable interaction effects. In addition to interactions, individual variations in the number of branches in the HPS also affected QRS timing. While future models should account for these potential sources of variability, this study indicates that minor anatomical differences between a healthy patient's HPS and that of a generic model are unlikely to significantly impact model fidelity or clinical interpretation when both systems are physiologically normal.
\end{abstract}
\newpage
\section{Introduction}
\noindent Cardiovascular disease remains a leading cause of death globally, accounting for over 19 million deaths in 2021 \cite{heart_disease_leading_death}. Among these, cardiac arrhythmias—irregular heart rhythms caused by disruptions in the heart’s electrical system—are major contributors to morbidity and sudden cardiac death \cite{arrythmias_death}. Cardiac digital twins (CDTs) have emerged as a potentially transformative tool in cardiology, offering enhanced precision and personalization in cardiac treatment. These patient-specific models simulate cardiac physiology, enabling clinicians to optimize therapies, improve treatment development, and provide personalized care \cite{digital_twin_definition, device_testing}. The effectiveness of CDTs in modeling arrhythmias depends on their ability to accurately model cardiac electrophysiology. The electrocardiogram (ECG) plays a critical role in this by offering a noninvasive validation framework. It provides a clear measure of the heart's electrical activity that can be compared to model-derived outputs, allowing for model validation and parameter adjustment in patient-specific settings, making it a crucial part of CDT models \cite{first_sigma_b, ecg_digital_twin, ecg_digital_twin_2}. \\

\noindent A key component of the cardiac electrical system is the His-Purkinje System (HPS), a specialized conduction system responsible for the rapid, coordinated depolarization of the ventricles \cite{what_are_purkinje}. The HPS allows for efficient electrical signal propagation across the ventricular myocardium, ensuring synchronized left and right ventricular contractions and rhythmic depolarization that generates the QRS complex on ECGs \cite{role_of_HPS, role_of_HPS_2, purkinje_contraction_sync}. Dysfunctional HPS pathways—as seen in conditions like bundle branch block (BBB)—disrupt this coordinated activity, producing characteristic QRS abnormalities that serve as diagnostic indicators of underlying pathology \cite{HPS_QRS_disease, Diagnosing_BBB, normal_ecg_and_disease_specs}.\\

\noindent Despite its critical role, current approaches to modeling the HPS for CDTs face a major limitation. Personalized geometries in CDTs are typically derived from clinical scans, such as MRIs or CT scans; however, the branches that make up the HPS are too thin to be visualized in these scans, forcing models to rely on assumptions about its structure \cite{scans_for_geometries, cant_see_HPS}. A non-invasive approach for generating patient-specific HPSs is using a fractal tree rule-based method to construct the Purkinje system on a patient-specific heart mesh, but this can introduce discrepancies between the model and actual patient anatomy based on parameters used in the algorithm \cite{fractal_tree_citation, fractal_tree_methods, using_fractal}.
\\

\noindent Sensitivity analysis (SA) is a critical approach to systematically evaluate how these minor discrepancies and variations in HPS properties affect system behavior and identify which input parameters drive variability in results. While many SAs have been done on ECGs in computational models, most have predominantly focused on electrophysiological properties (e.g., ion channel kinetics) or ECG specific parameters, such as lead placement \cite{example_SA_ECG_OAT, example_SA_ECG_OAT_2, example_SA_ECG_OAT_3}. Despite the critical role of the HPS in shaping ventricular depolarization, the influence of its structural and physical properties—such as number of branches or branching angles—on QRS complex morphology remains understudied \cite{second_sigma_b}. Previous studies examining the impact of HPS structure on model outputs have not fully accounted for the complete HPS structure or the interactions between its structural features, which can contribute to variability \cite{previous_HPS_SA}. Addressing this gap is essential to understanding the influence of HPS assumptions on ECG morphology and refine the accuracy of CDTs \cite{including_HPS_CDT}. \\

\noindent This paper presents a global, variance-based sensitivity analysis to evaluate the impact of minor variations in the HPS's physical structure on QRS complex morphology. Unlike traditional local sensitivity approaches, a global sensitivity approach enables the examination of both the individual effects of structural parameters and their interactions \cite{Pras_GSA_1}. We vary nine selected parameters that govern the personalized anatomies of HPS structures and quantify the resulting variability in ten quantities of interest (QOIs) related to QRS morphology. We compute Sobol sensitivity indices to identify which HPS parameters drive the variability observed in the QOI distributions, not only through direct effects but also through their interactions with other HPS parameters. Our findings highlight key factors of the HPS structure that influence variability in simulated ECG QRS complexes, offering valuable insights into CDT fidelity. This work bridges computational and clinical frameworks, enhancing the reliability of CDTs in arrhythmia management and personalized cardiology.
\section{Methods}

\noindent This study aims to quantify the variability across 10 QOIs relating to QRS morphology resulting from HPS parameter variations and to identify the parameters that drive this variability. First, variation ranges are assigned to the HPS parameters, and a Saltelli sampling scheme is used to simultaneously vary all parameters across different trials. Using the sampled parameter values, Purkinje networks are generated and incorporated into a heart simulation to produce ECG signals. The resulting ECG signals are then analyzed to extract the 10 QOIs from QRS complexes. Once all sampled Purkinje networks have been simulated and their corresponding QOIs identified, the distributions of the QOIs are visualized using box plots and their variability is quantified. Finally, a Sobol sensitivity analysis is conducted to determine which HPS parameters contribute most to the observed variability. 

\subsection{Generating Purkinje Fiber Networks}
\label{sec:generate_purkinkje}
\begin{figure}[h!]
\centering
\includegraphics[width=0.75\textwidth]{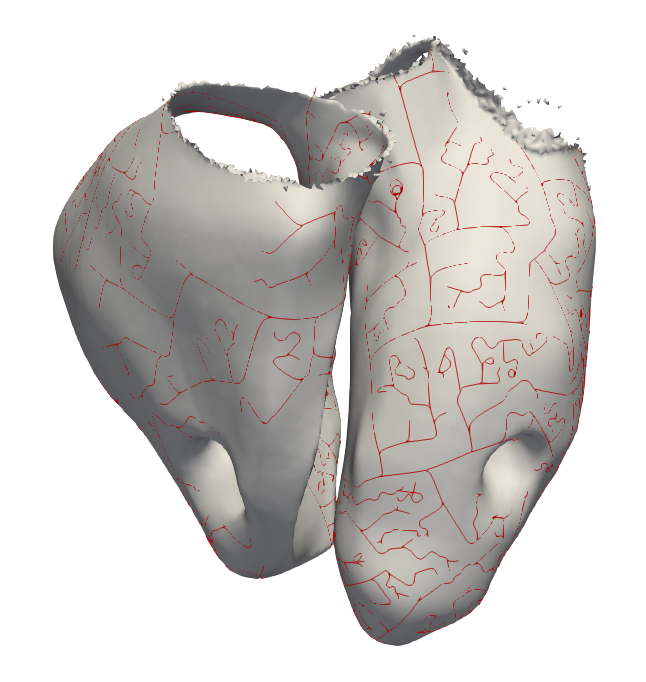} 
\caption{Purkinje fiber network generated using \textit{fractal-tree}. The network is visualized on a bi-ventricular model to illustrate its branching pattern and spatial distribution across the left and right ventricular surfaces.}
\label{fig:purkinje_on_vent} 
\end{figure}

\noindent Purkinje fiber networks used in this study were generated using \textit{fractal-tree}, an open-source Python software for generating Purkinje systems \cite{fractal_tree_citation}. This algorithm constructs networks iteratively, starting with an initial node and direction. Subsequent branches are generated through a randomized process governed by the parameters number of branches, minimum branch length, median branch length, branch length standard deviation, branch angle, branch repulsivity,  fascicle angles, and fascicle length. While these parameters collectively define the overall network structure, the inherent randomness of the branching process introduces slight variations between individual networks. After each branch is generated, the algorithm projects the newly created node back onto the input mesh surface, ensuring accurate growth of the network on a non-smooth geometry.\\

\noindent To restrict the growth of the Purkinje system to the ventricles, the surfaces of the left and right ventricles were extracted from the whole heart mesh, ensuring that the septum was excluded (see Figure \ref{fig:purkinje_on_vent}). All networks in this study were generated on this bi-ventricular model.

\subsection{Computing ECG signals}

\noindent The heart mesh used in this study was constructed from de-identified cardiac CT images provided by Siemens Healthineers. Further detail on the anatomy of the mesh is provided in \cite{davey_fsi_simulation}. Electrical propagation through the heart was modeled using a monodomain reaction-diffusion model \cite{Monodomain_model, LaryssaThesis}. The electrophysiology of cardiomyocytes in the atria, Purkinje fibers, and ventricles was modeled using the Nash-Panfilov ionic model \cite{nash_panfilov_model}. Ionic dynamics were integrated using a second-order strong-stability preserving Runge-Kutta method \cite{ssprk2}. \\

\noindent ECG signals were computed by recovering the extracellular potentials from the monodomain equation. Let \(\phi_\text{e}\) denote the extracellular potential. Assuming the heart is situated within an unbounded volume conductor, \(\phi_\text{e}\) was computed in a standard monodomain simulation via \cite{ecg_eq_1}:

\begin{equation}
    \phi_e = \frac{1}{4 \pi \sigma_\text{b}} \int_{\Omega} \frac{\chi I_\text{m}}{||\mathbf{r}||} \, d\Omega
\end{equation}

\noindent Here, \(\sigma_\text{b}\) represents the conductivity tensor of the volume conductor (the torso conductivity tensor), \(I_\text{m}\) is the total transmembrane current, and \(\mathbf{r}\) is the distance vector from the field points of the heart mesh to the electrode. $\sigma_\text{b}$ was set to $2.16  \frac{\text{mS}}{\text{cm}}$ to align with heart model simulations conducted in previous studies that were based on patient data \cite{first_sigma_b, second_sigma_b}. The source term \(\chi I_\text{m}\) was computed using \cite{ecg_eq_2}:

\begin{equation}
    \chi I_\text{m} = \nabla \cdot (\sigma_\text{m} \nabla V_\text{m})
\end{equation}

\noindent which was computed using values obtained from the monodomain model. Extracellular potential calculations were verified by replicating benchmark presented in \cite{ecg_eq_1}. Supplementary Methods Section \nameref{sec:ecg_validation} presents this validation study and results. This validation step confirmed that the recovered ECG traces were computed correctly and exhibited the anticipated features. Raw extracellular potential signals were processed using a low-pass filter from the open-source library ECG-Deli, with a cutoff of 10 Hz \cite{ECGDeli}.\\

\begin{figure}[h!]
\centering
\includegraphics[width=0.75\textwidth]{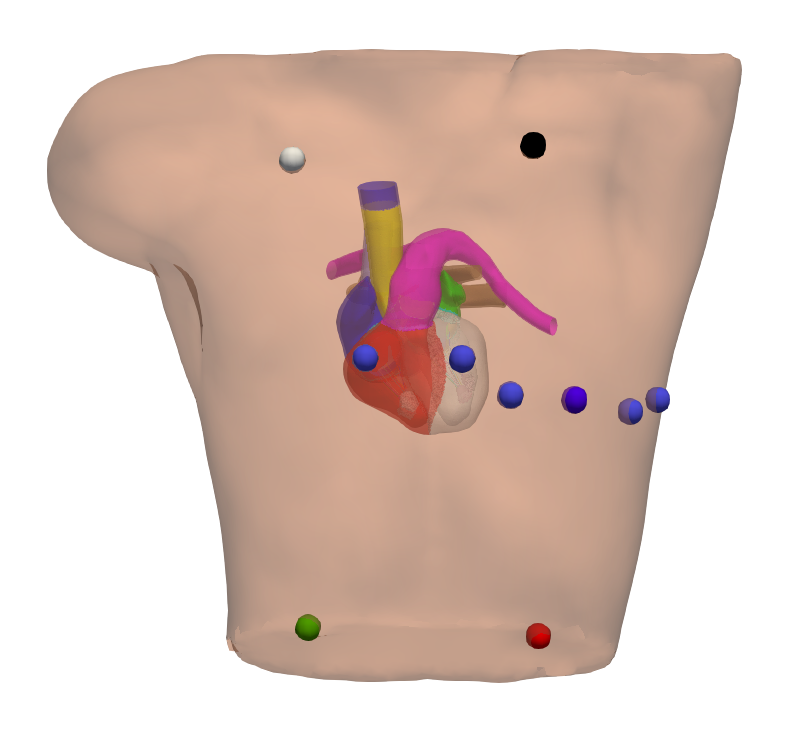} 
\caption{Whole heart model with overlaid Purkinje fiber system, embedded in an open-source torso model to determine spatial electrode distribution for 12-lead ECG.}
\label{fig:12_lead_torso} 
\end{figure}

\noindent To determine electrode locations, the heart model was placed within an open source torso model, which was modified to reflect anatomically accurate dimensions of an adult male, with a torso height of 46.5 cm and a heart depth of 3.75 cm \cite{torso_model, torso_dimensions}. Electrodes were then placed according to the standard 12-lead ECG configuration, see Figure \ref{fig:12_lead_torso} \cite{ecg_lead_placement}.  \\

\noindent In this study, extracellular potential measurements were recorded in the precordial leads V2, V4, and V6, as well as the left augmented vector lead (aVL). The aVL lead was derived using the following formula \cite{aVL_formula}:

\begin{equation*}
    \phi_\text{aVL} = \phi_\text{LA} - \frac{\phi_\text{RA} + \phi_\text{LL}}{2}
\end{equation*}

\noindent where $\phi_\text{LA}$, $\phi_\text{RA}$, and $\phi_\text{LL}$ represent the potentials computed by Equations (2) and (3) at the left arm, right arm, and left leg electrodes, respectively. \\

\noindent All trial simulations were performed on a high-performance computing cluster consisting of over 14,000 computing cores across 78 nodes. Each simulation utilized five cores and solved the monodomain equation at a time step of 0.0156 ms over a total duration of 325 ms. Extracellular potentials were recorded at every fifth monodomain time-step (every 0.078 ms). A first-order Implicit-Explicit backward-differentiation scheme was employed as the numerical time integrator \cite{IMEX}. 

\subsection{Nominal Simulation and Parameter Sampling}

\noindent A nominal simulation was conducted to serve as the baseline for parameter ranges. The parameters used to generate the Purkinje fiber system in this simulation are defined in Table \ref{Nominal_parameters_table}. These values were chosen to generate a QRS complex with normal morphological features for an adult male based on guidelines presented from standardized ECG interpretation guidelines, see Table \ref{QOI_Nominal} \cite{normal_ecg_and_disease_specs}.\\

\begin{table}[h!]
    \centering
\caption{Nominal parameter values used in the initial simulation.}
    \vspace{5 mm}
    \setlength{\tabcolsep}{20pt} 
    \begin{tabular}{@{}llc@{}}
        \toprule
        \textbf{Parameter}                 & \textbf{Value} \\
        \midrule
        Initial branch length                   & 50 mm                 \\
        Number of branch generations           & 13                    \\
        Median branch length                   & 6 mm                   \\
        Length of segments in branch            & 0.1 mm\\
        Branch angle                           & 0.2 rad        \\
        LV Fascicle Lengths             & [5,5] mm         \\
        LV Fascicle Angles              & [-0.1, 0.2] rad  \\
        Repulsion Parameter                    & 0.1 \\
        \bottomrule
    \end{tabular}
    \label{Nominal_parameters_table}
\end{table}

\noindent In all simulations, the Purkinje network of the left ventricle comprised of two fascicles, with the parameters of each fascicle treated as independent variables in the SA. To ensure alignment between the Purkinje network and the heart mesh, the initial branch length ($B_\text{L}$) was fixed at 50 mm in all simulations. To ensure non-negative and proportionally sized branches in trial networks, the standard deviation of branch length was set to Median Branch Length ($M_{B_\text{L}}$) times by square root of 2, and the minimum branch length was fixed at \( \frac{M_{B_\text{L}}}{10} \) for all simulations. The number of branch generations was rounded to the nearest integer in all simulations and varied within a range of 9 to 17. The remaining parameters were assigned a variation range of ±30\% from their nominal values. Since little is known about the true distribution of HPS structure among human populations, this variation range was chosen to mimic moderately large differences in HPS structure. Samples were generated using a Saltelli sampling scheme, with 2048 points per dimension, resulting in a total of 22,528 simulations \cite{saltelli_sampling}. The number of samples was selected to balance minimizing error and reducing the computational cost of running numerous simulations.\\

\begin{figure}[ht]
\centering
\includegraphics[width=1.0\textwidth]{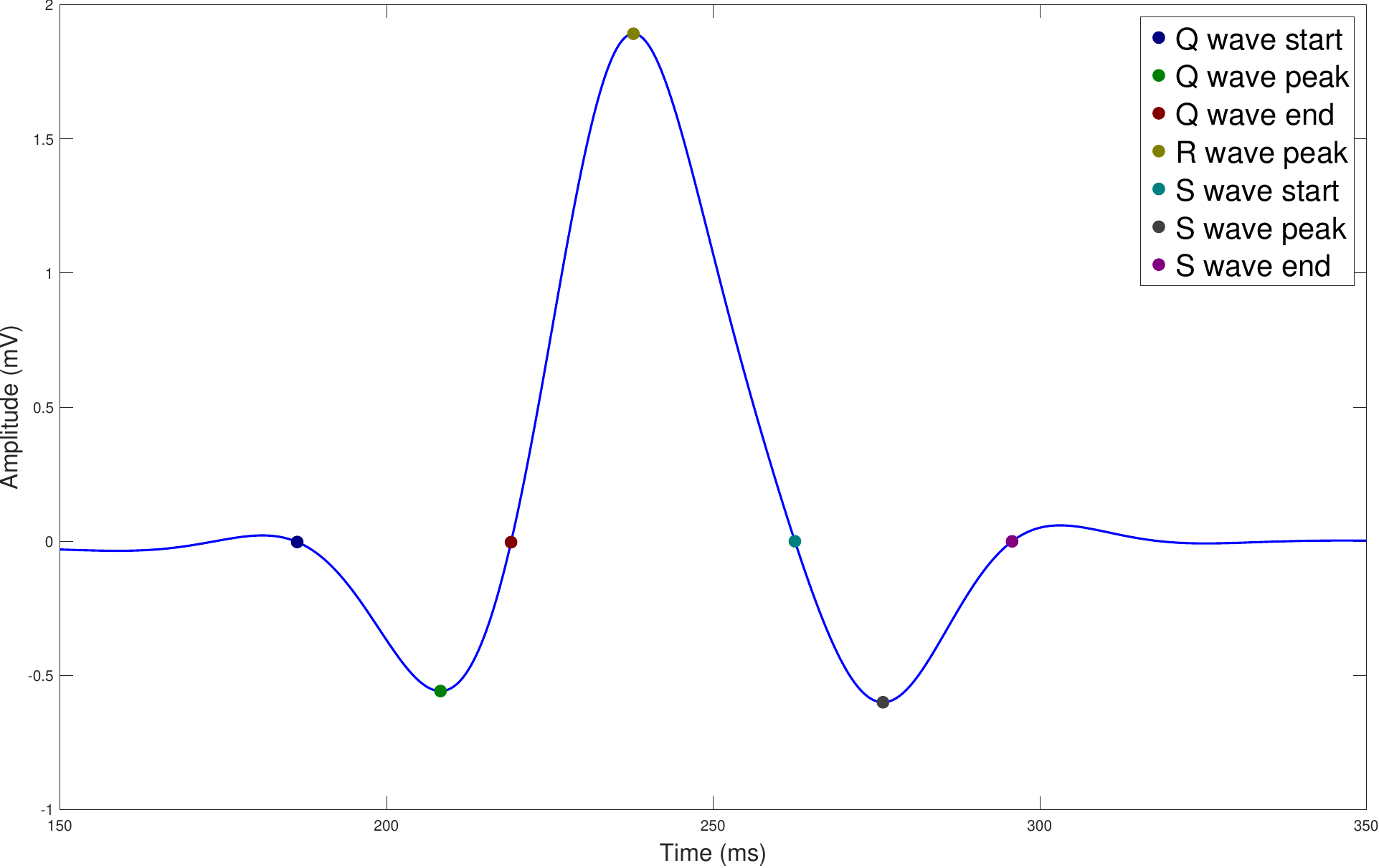} 
\caption{QRS complex of an ECG trace recorded at V2 during nominal simulation. Demonstrates MATLAB algorithm for identification of Q, R, and S wave.}
\label{fig:nominal_QRS_COmplex} 
\end{figure}

\noindent An algorithm implemented in MATLAB was used to extract the start, peak, and end times, as well as the amplitudes, of the Q, R, and S waves from the filtered signal. As this analysis focuses exclusively on how the HPS influences ventricular depolarization, P and T waves were not characterized in this study. Figure \ref{fig:nominal_QRS_COmplex} illustrates the identification of peaks in the filtered signal recorded at electrode V2, generated from the initial simulation using the nominal parameters.\\

\noindent From the identified locations, the following quantities of interest were computed and subsequently used in the sensitivity analysis: Q-wave duration, peak time, and peak amplitude; R-wave duration, peak time, and peak amplitude; S-wave duration, peak time, and peak amplitude; and QRS duration. Table \ref{QOI_Nominal} presents the QOI for the V2 QRS complex of the nominal simulation.

\subsection{Data Analysis and Sensitivity Indices}

\noindent To visualize the distributions of each QOI, box plots were generated for both the original and standardized z-score distributions using SAS statistical software. z-scores were used to allow for unitless comparisons between the different QOIs. Statistical measures—including the mean ($\mu$), standard deviation ($\sigma$), Coefficient of Variation (CV), z-score first and third quartiles ($z_{Q1}$ and $z_{Q3}$), and z-score range ($z_{\text{range}}$)—were computed using SAS. These values were used to quantify the variability observed in the QOI distributions caused by variations in HPS parameters.

\subsubsection{Sobol Sensitivity Analysis}
\label{sec:sobol_calc}

\noindent In addition to quantifying variability, the main and total Sobol sensitivity indices were computed to identify which HPS parameters contribute to the observed variability \cite{sobol_indices_1}. The main effect, also known as the first-order Sobol index ($S_1$), represents the proportion of output variance attributed solely to a single parameter, excluding interactions with other parameters.\\

\noindent Let \( X_i \) be a parameter of the model and \( Y = f(X) \) be the output of interest over the parameter space \( X \). The total variance of the output is given by $\mathbb{V}(Y) = \mathbb{E}(Y^2) - (\mathbb{E}(Y))^2$, in which \( \mathbb{E}(Y) = \int Y P(X) \, dX \) and \( P(X) \) is the probability density function of \( X \). The first-order Sobol sensitivity index is then calculated as \cite{Pras_GSA_1, sobol_indices_2}:

\begin{equation*}
    S_i = \frac{\mathbb{V}(\mathbb{E}(Y|X_i))}{\mathbb{V}(Y)} = \frac{\mathbb{V}_i(Y)}{\mathbb{V}(Y)}
\end{equation*}

\noindent The total Sobol index ($S_\text{T}$) captures both the direct contribution of the parameter to the output variance ($S_1$) and the contribution of the parameters interactions with other parameters in driving output variability. It is calculated as the total variance of the output associated with the parameter, including variances caused by interactions, divided by the total variance of the output. Therefore, it represents the sum of the main index ($S_1$) and the higher-order interaction indices ($S_2$,$S_3$,...) for the parameter \cite{sobol_indices_2}.\\

\noindent For both $S_1$ and $S_\text{T}$, significant indices were considered those that were greater than 0.05, meaning that the parameter accounts for greater than 5\% of the variability observed in the QOIs distribution \cite{sobol_interpretation_guide}. The first-order and total Sobol indices were calculated using the open-source Python library SALib \cite{SALIB_1, SALIB_2}.

\section{Results}

\noindent The results from the 22,528 trials conducted in this study are presented below. Overall, variability across all QOIs was minimal. Nearly all distributions appeared visually symmetric, with only a small number of outliers. In this study, outliers are defined as values falling outside the range $[-1.5 \times z_{\text{IQR}}, 1.5 \times z_{\text{IQR}}]$. Figures showing both the original and standardized distributions are provided in the \nameref{sec:QOI_distributions} Section of Supplementary Materials. \\

\subsection{Peak Amplitudes and Wave Durations}
\begin{table}[h!]
    \centering
    \caption{Summary Statistics for Peak Amplitudes (mV) across 4 Leads}
    {\fontsize{9}{11}\selectfont 
    \renewcommand{\arraystretch}{1.45} 
    \setlength{\tabcolsep}{3.5pt} 
    \begin{tabular}{l *{9}{c}} 
        \toprule
        & \multicolumn{3}{c}{Q Wave} 
        & \multicolumn{3}{c}{R Wave} 
        & \multicolumn{3}{c}{S Wave} \\
        \cmidrule(lr){2-4} \cmidrule(lr){5-7} \cmidrule(lr){8-10}
        & Mean ($\sigma$) & CV (\%) & $z_{\text{range}}$ 
        & Mean ($\sigma$) & CV (\%) & $z_{\text{range}}$  
        & Mean ($\sigma$) & CV (\%) & $z_{\text{range}}$  \\
        \midrule
        V2 & -0.493 (0.0760) & -15.42 & 6.74
          & 2.03 (0.183) & 9.03 & 10.186 
          & -0.584 (0.180) & -30.83 & 5.97 \\
        V4 & -0.315 (0.0449) & -14.23 & 7.58 
          & 1.21 (0.0657) & 5.42 & 17.93
          & -0.227 (0.0811) & -35.72 & 6.20 \\
        V6 & -0.223 (0.0306) & -13.70 & 7.92 
          & 0.763 (0.0420) & 5.50 & 17.99 
          & -0.107 (0.0410) & -38.27 & 6.32 \\
        aVL & -0.0171 (0.00349) & -20.41 & 5.84
          & 0.147 (0.0164) & 11.10 & 9.04 
          & -0.0229 (0.00700) &  -30.53 & 11.85 \\
        \bottomrule
    \end{tabular}
    }
    \label{tab:stats_amplitude}
\end{table} 
\begin{table}[h!]
    \centering
    \caption{Summary Statistics for Wave Durations (ms) across 4 Leads}
    {\fontsize{9}{11}\selectfont 
    \renewcommand{\arraystretch}{1.45} 
    \setlength{\tabcolsep}{3.5pt} 
    \begin{tabular}{l *{9}{c}} 
        \toprule
        & \multicolumn{3}{c}{Q Wave} 
        & \multicolumn{3}{c}{R Wave} 
        & \multicolumn{3}{c}{S Wave} \\
        \cmidrule(lr){2-4} \cmidrule(lr){5-7} \cmidrule(lr){8-10}
        & Mean ($\sigma$) & CV (\%) & $z_{\text{range}}$ 
        & Mean ($\sigma$) & CV (\%) & $z_{\text{range}}$  
        & Mean ($\sigma$) & CV (\%) & $z_{\text{range}}$  \\
        \midrule
        V2 & 22.79 (3.38) & 14.84 & 7.25 
          & 45.04 (2.91) & 6.46 & 8.20 
          & 27.27 (2.50) & 9.16  & 13.41 \\
        V4 & 23.40 (2.84) & 12.13 & 7.82 
          & 49.59 (2.94) & 5.92 & 5.96 
          & 22.06 (2.31) & 10.48 & 8.92 \\
        V6 & 23.64 (2.68) & 11.35 & 7.60
          & 49.15 (3.12) & 6.35 & 7.34
          & 20.88 (2.68) & 12.84 & 9.34 \\
        aVL & 18.68 (2.85) & 15.23 & 7.17 
          & 50.09 (1.78) & 3.56 & 11.80
          & 20.57 (2.66) & 12.96 & 12.78 \\
        \bottomrule
    \end{tabular}
    }
    \label{tab:stats_duration}
\end{table}

\noindent Tables~\ref{tab:stats_amplitude} and \ref{tab:stats_duration} summarize the statistics for individual wave duration and peak amplitude distributions. The standard deviations for these QOI are generally low, with the greatest variability observed in the R-wave peak amplitude of lead V2 (\(\sigma = 0.183\) mV) and the Q-wave duration of lead V2 (\(\sigma = 3.38\) ms). The $\text{CV}$ values are generally low across these QOI. Though the peak amplitude CV values are particularly high, this is due to the very small and sometimes negative \(\mu\) values, a characteristic that is known to inflate the CV. The moderate CV values observed in Q wave duration in lead V2 ($\approx 15\%$) is likely driven by extreme outliers in this distribution. \\

\noindent The interquartile range of z-scores was narrow, with $z_{\text{Q1}} > -0.75$ and $z_{\text{Q3}} < 0.75$, indicating that the majority of data points lie within approximately 0.75 standard deviations of the mean. Standardized quartiles for all QOI are presented in \nameref{sec:standardized_quartiles} section of Supplementary Materials.  When excluding outliers, nearly all data points fell within 2.5 standard deviations of the mean, with $z_{\text{range}} \lesssim 5$ across all QOI distributions. The number of outlier trials for these QOI was relatively small compared to the total number of trials conducted was very low (<1\%) and most clustered near the boundaries of the whisker plots, suggesting that they only mildly deviate from the bulk of the data. \\

\noindent Although the majority of outliers were located near these boundaries, a few trials exhibited more extreme deviations, which contributed to inflated measures of variability. When such outliers were included in the analysis, $z_{\text{range}}$ values increased substantially, ranging from 5.84 to 17.99 for peak amplitudes and from 5.96 to 13.41 for wave durations. One specific trial was identified as a significant outlier across multiple QOI distributions. This trial was responsible for the pronounced inflation in $z_{\text{range}}$ values observed in the S-wave peak amplitude of lead aVL, the R-wave peak amplitude of leads V4 and V6, the S-wave duration across all leads, and the R-wave duration in lead aVL. This outlier trial is visualized in Figure~\ref{fig:outlier_trial} in the Supplementary Figures.

\begin{table}[h!]
    \centering
    \caption{Summary Statistics for QRS wave duration (ms) across 4 Leads}
    {\fontsize{11}{11}\selectfont 
    \renewcommand{\arraystretch}{1.45} 
    \setlength{\tabcolsep}{4.5pt} 
    \begin{tabular}{l c c c} 
        \toprule
        Lead & Mean ($\sigma$) & CV (\%) & $z_{range}$  \\
        \midrule
        V2  & 95.10 (3.88)& 4.08 & 7.19 \\
        V4  & 95.05 (3.77)& 3.97  & 7.04 \\
        V6  & 93.67 (3.69)& 3.94  & 7.06 \\
        aVL & 89.33 (2.03)& 2.28  & 11.68 \\
        \bottomrule
    \end{tabular}
    }
    \label{tab:stats_QRS_duration}
\end{table}

\noindent Table \ref{tab:stats_QRS_duration} present the summary statistics for QRS complex duration distributions. Similar to individual wave durations, the overall QRS complex duration exhibited low variability, with the highest standard deviation observed in lead V2 ($\sigma = 3.88$~ms). $\text{CV}$ values were also extremely low for these distributions. The lower $\sigma$ and $\text{CV}$ values for these distributions compared to individual wave distributions are likely attributable to compensatory effects among the individual wave duration distributions, where broader spreads in some waveforms are offset by narrower spreads in others. \\

\noindent $z_{\text{IQR}}$ values for these distributions were similar to those of individual wave durations, with $z_{\text{Q1}} > -0.7$ and $z_{\text{Q3}} < 0.7$, indicating that the majority of data points lie within approximately 0.7 standard deviations of the mean. Furthermore, the $z_{\text{range}}$ values—when outliers were excluded—for QRS complex duration were also similar to those of individual wave durations, with nearly all data points falling within 2.5 standard deviations of the mean, $z_{\text{range}} \lesssim 5$ across all QOI distributions.\\

\noindent The proportion of outlier trials relative to the total number of trials was low, with most outliers clustering near the boundaries of the whisker plots. However, a few extreme outlier trials were present, resulting in elevated $z_{\text{range}}$ values. Specifically, $z_{\text{range}}$ for overall QRS duration distributions ranged from 7.04 to 11.68. The previously identified outlier trial was also a significant anomaly in the QRS complex of lead aVL, contributing to the maximum observed $z_{\text{range}}$ value of 11.68 for this QOI.

\subsection{Peak Times}

\begin{table}[h!]
    \centering
    \caption{Summary Statistics for Peak Times (ms) across 4 Leads}
    {\fontsize{9}{11}\selectfont 
    \renewcommand{\arraystretch}{1.45} 
    \setlength{\tabcolsep}{3.5pt} 
    \begin{tabular}{l *{9}{c}} 
        \toprule
        & \multicolumn{3}{c}{Q Wave} 
        & \multicolumn{3}{c}{R Wave} 
        & \multicolumn{3}{c}{S Wave} \\
        \cmidrule(lr){2-4} \cmidrule(lr){5-7} \cmidrule(lr){8-10}
        & Mean ($\sigma$) & CV (\%) & $z_{\text{range}}$ 
        & Mean ($\sigma$) & CV (\%) & $z_{\text{range}}$  
        & Mean ($\sigma$) & CV (\%) & $z_{\text{range}}$  \\
        \midrule
        V2 & 213.98 (6.06) & 2.83 & 29.14
          & 240.16 (5.78) & 2.41 & 30.24 
          & 279.32 (5.90) & 2.11 & 30.08 \\
        V4 & 214.81 (6.19) & 2.88 & 28.58 
          & 242.046(6.29) & 2.60 & 27.80 
          & 282.19 (5.95) & 2.11 & 29.61 \\
        V6 & 214.99 (6.24) & 2.90 & 28.35 
          & 241.81 (6.35) & 2.63 & 27.54 
          & 281.92 (5.99) & 2.13 & 29.45 \\
        aVL & 214.07 (6.02) & 2.81 & 29.28 
          & 241.77 (5.89) & 2.43 & 29.75 
          & 279.24 (5.99) & 2.15 & 29.38 \\
        \bottomrule
    \end{tabular}
    }
    \label{tab:stats_wave_time}
\end{table}

\begin{figure}
    \centering
    \includegraphics[width=0.75\linewidth]{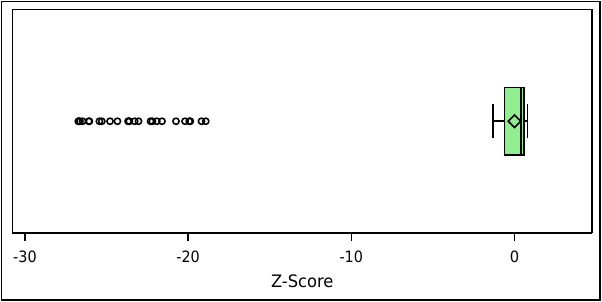}
    \caption{Distribution of R Wave peak times in lead V6}
    \label{fig:times_rep_figure}
\end{figure}

\begin{figure}
    \centering
    \begin{tabular}{cc}
        \includegraphics[width=0.5\textwidth]{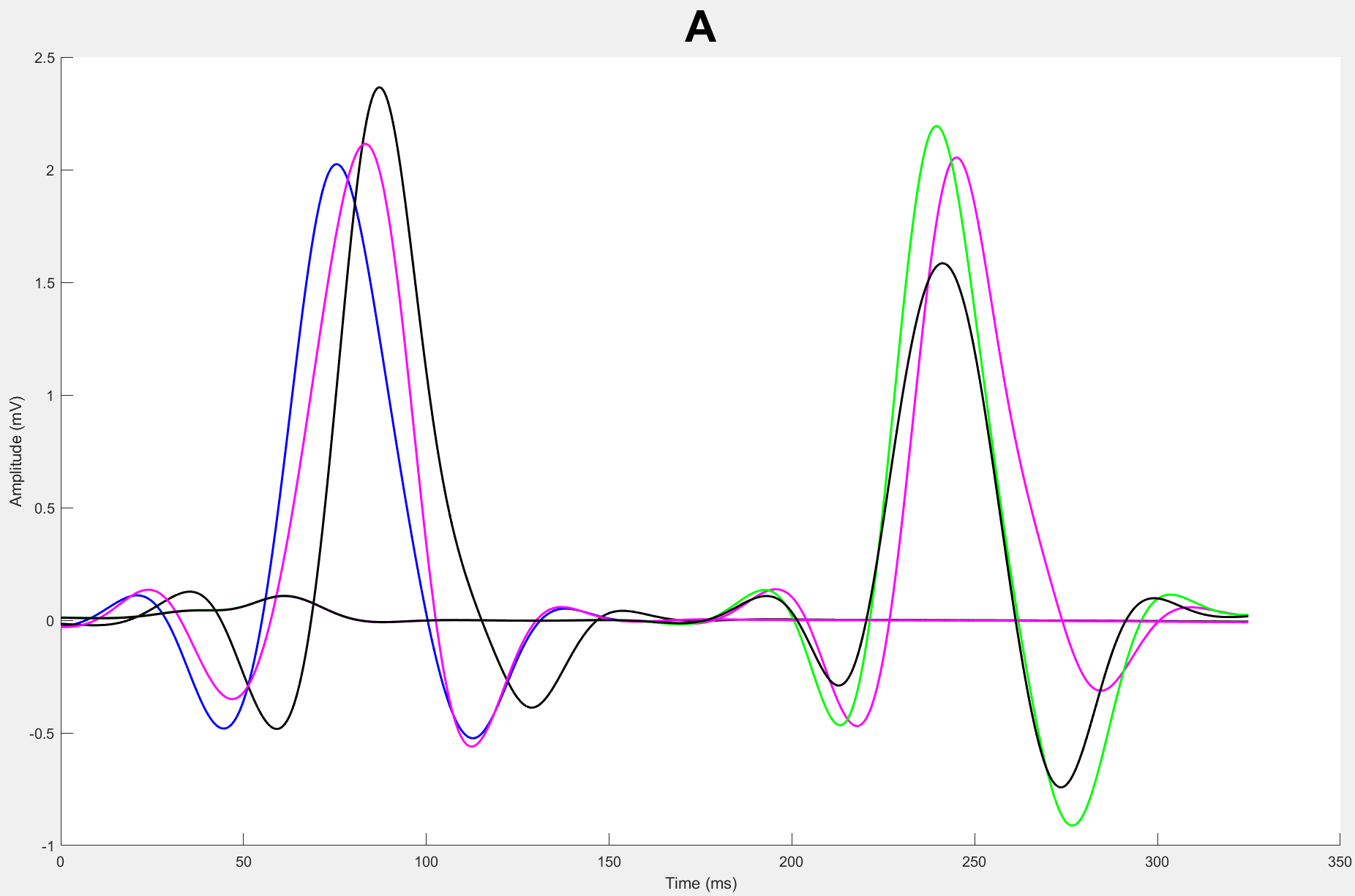} & \includegraphics[width=0.5\textwidth]{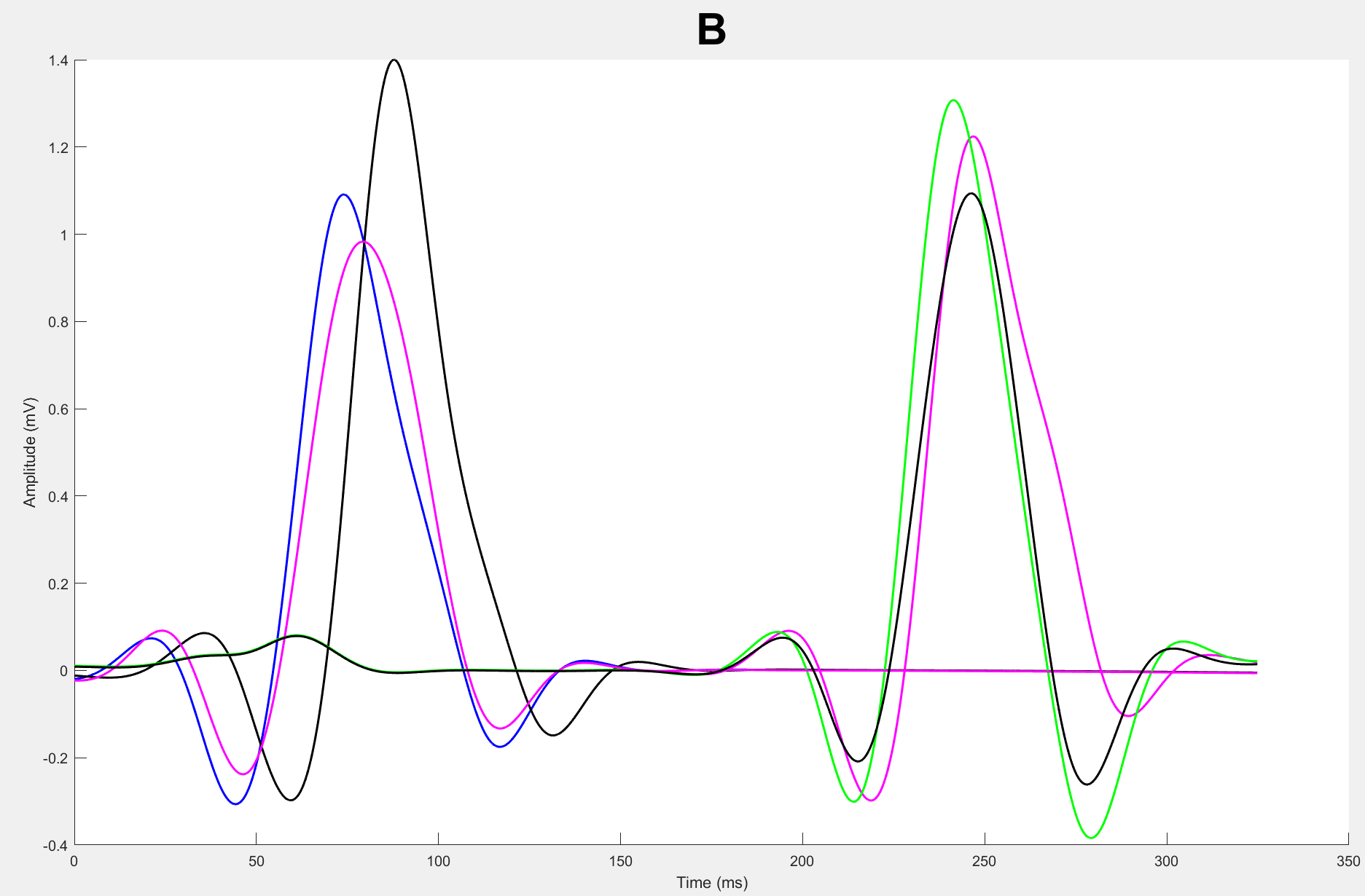} \\
        \includegraphics[width=0.5\textwidth]{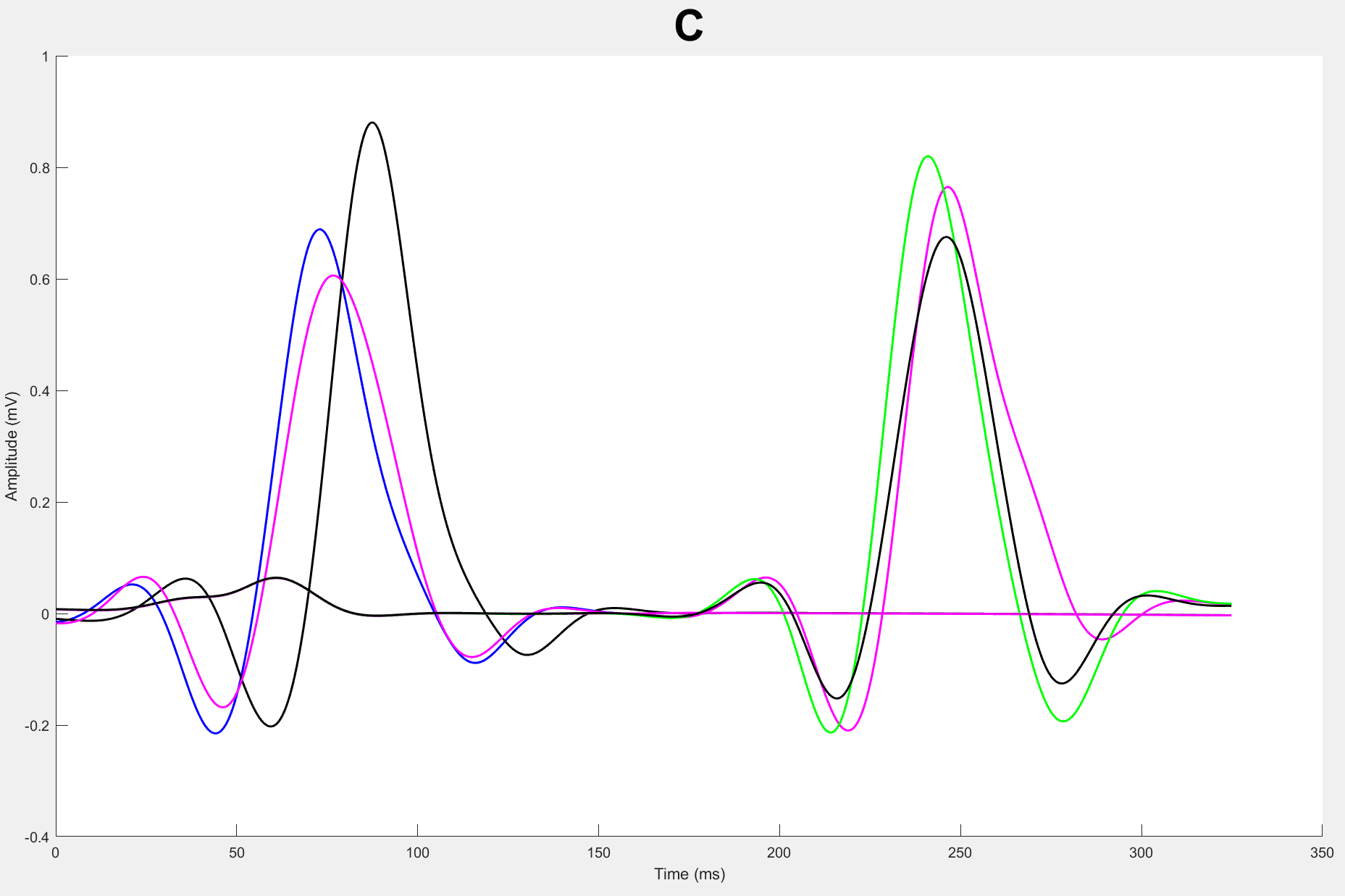} & \includegraphics[width=0.5\textwidth]{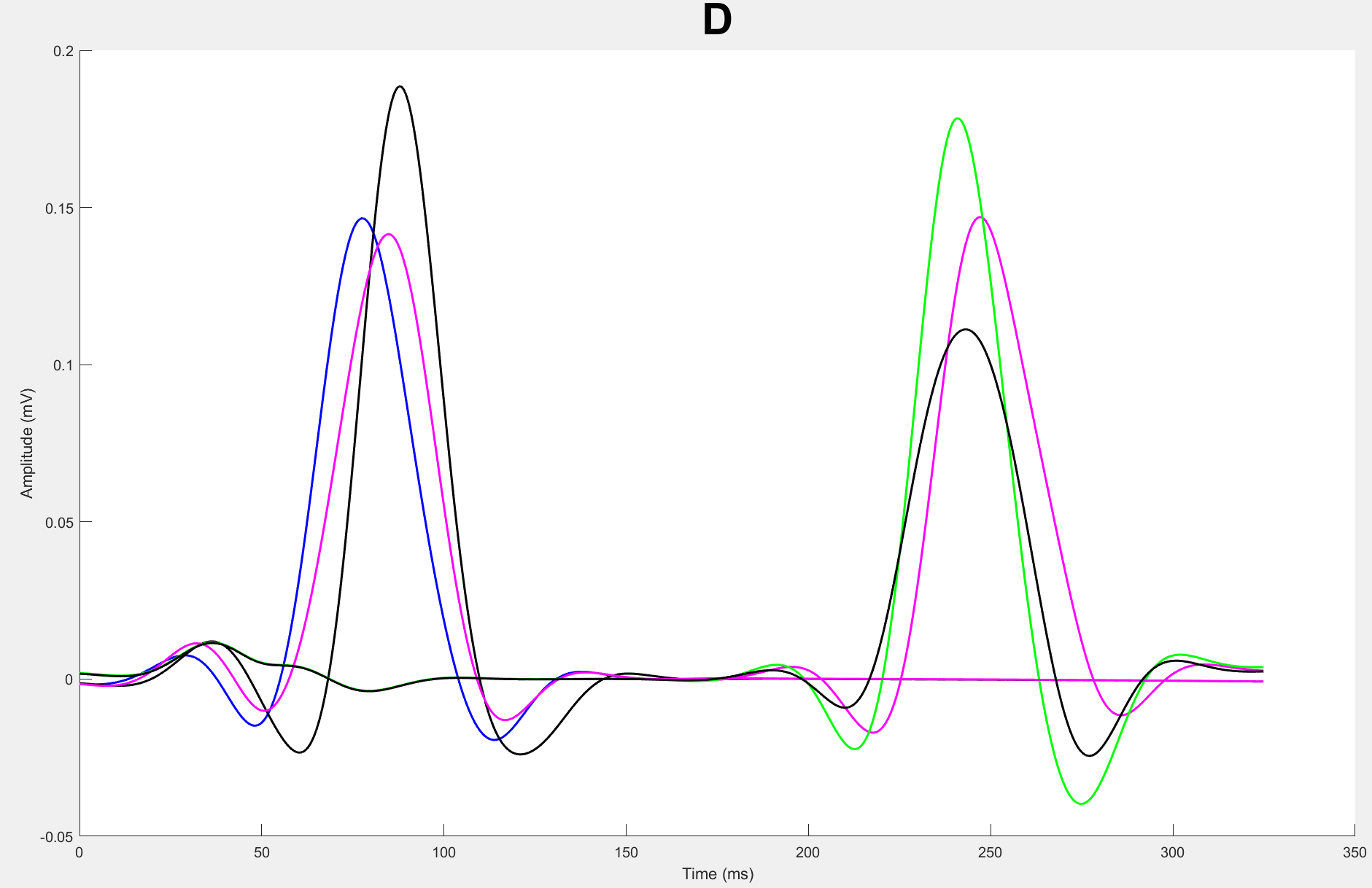} \\
    \end{tabular}
    \caption{Six trial QRS complexes, with three being outliers clustered around the second central point and three centered at the median, for V2 (A), V4 (B), V6 (C), and aVL (D).}
    \label{fig:supp_image_2}
\end{figure}

\noindent Table~\ref{tab:stats_QRS_duration} presents the summary statistics for the peak time distributions. Compared to the previously discussed QOIs, peak times exhibit a significantly greater spread, likely due to the distribution of its outliers. $z_{\text{IQR}}$ and $z_{\text{range}}$ values ignoring outliers for peak times was similar to those of the previously discussed QOI, with z-score quartiles within 0.7 standard deviations of the mean and nearly all data falling within 2.5 standard deviations of the mean. A representative distribution of the peak times is presented in Figure~\ref{fig:times_rep_figure}. This distribution illustrates how the outliers for peak times distributions are distributed clustering away from the whisker boundaries at a second central point. All peak time distributions displayed the same pattern, with the same outliers trials clustering away from the whisker bounds and at this point. To further investigate this, six representative trials were selected: three outliers clustered around the secondary central point and three data points aligned more closely with the primary distribution (see Figure~\ref{fig:supp_image_2}). This analysis revealed that at this secondary point, complete QRS complexes formed, suggesting that specific combinations of HPS parameters accelerate QRS complex formation. \\

 \noindent This phenomenon has caused substantial inflation of $z_{\text{range}}$ values across all peak time distributions, with values ranging from 27.54 to 30.24. Furthermore, this pattern also contributed to a higher $\sigma$ across all distributions. However, the overall variability remained low, with the highest standard deviation observed in the R-wave peak time of lead V6 (\(\sigma = 6.35\)). The $\text{CV}$ values were also extremely low for these distributions, with all values less than 3\%. This is likely due to the relatively low number of outlier compared to the total number of trials ran (0.11\% of trials were outliers). 
\subsection{Sobol Analysis}
\begin{figure}[h!]
    \centering
    \includegraphics[width=0.33\textwidth]{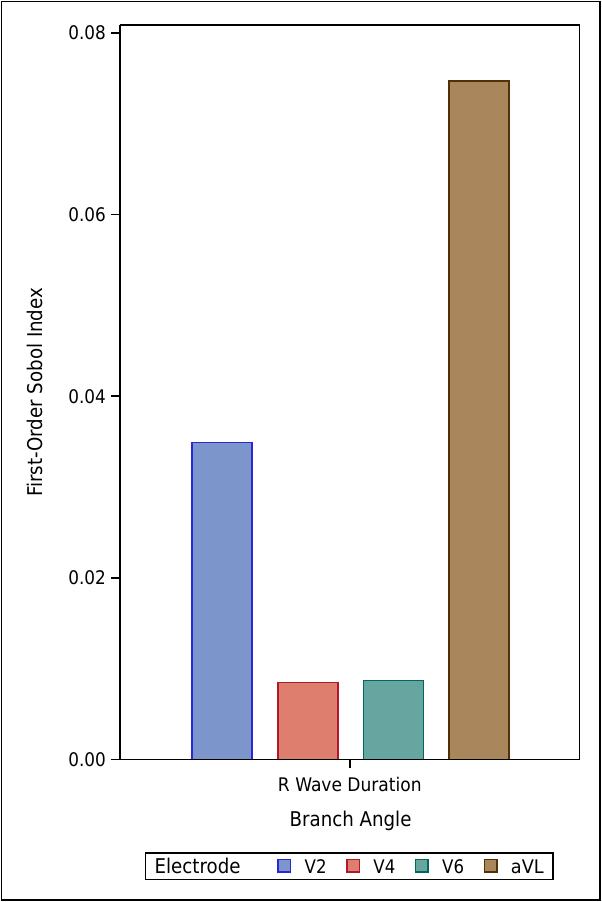}%
    \includegraphics[width=0.33\textwidth]{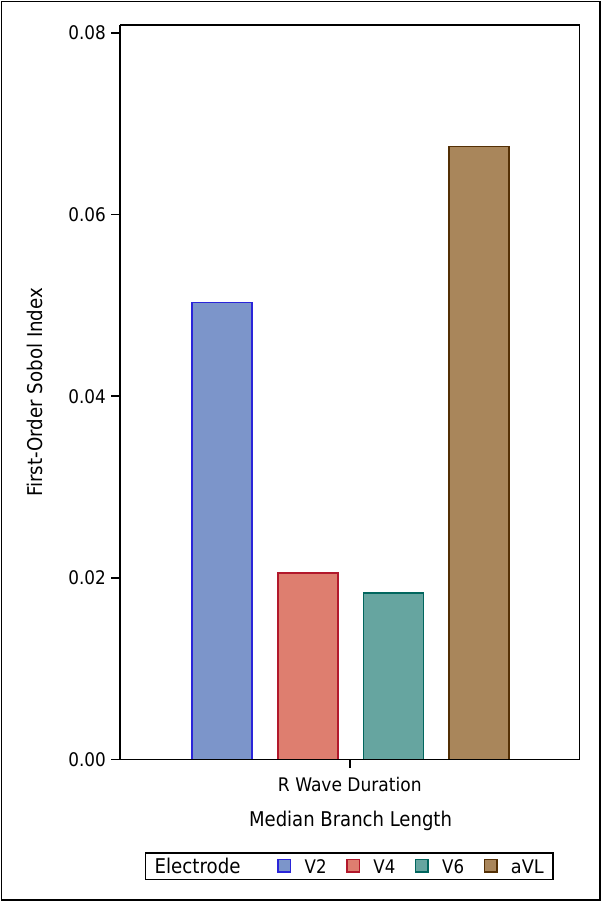}%
    \includegraphics[width=0.33\textwidth]{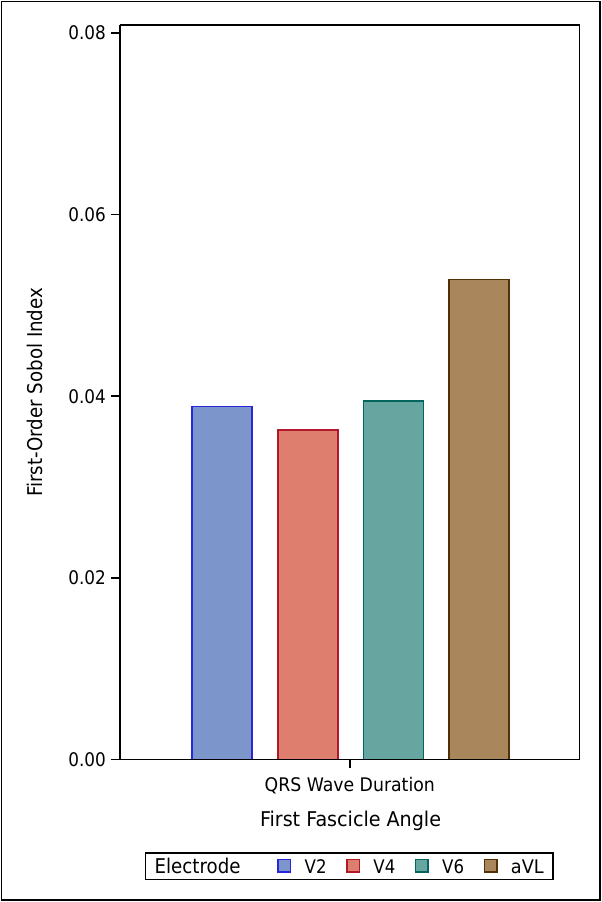}
    \caption{First-order Sobol indices for wave durations, highlighting cases where at least one lead shows a statistically significant contribution.}
    \label{fig:sig_wave_dur_sob} 
\end{figure}

\begin{figure}[h!]
    \centering
    \includegraphics[width=0.33\textwidth]{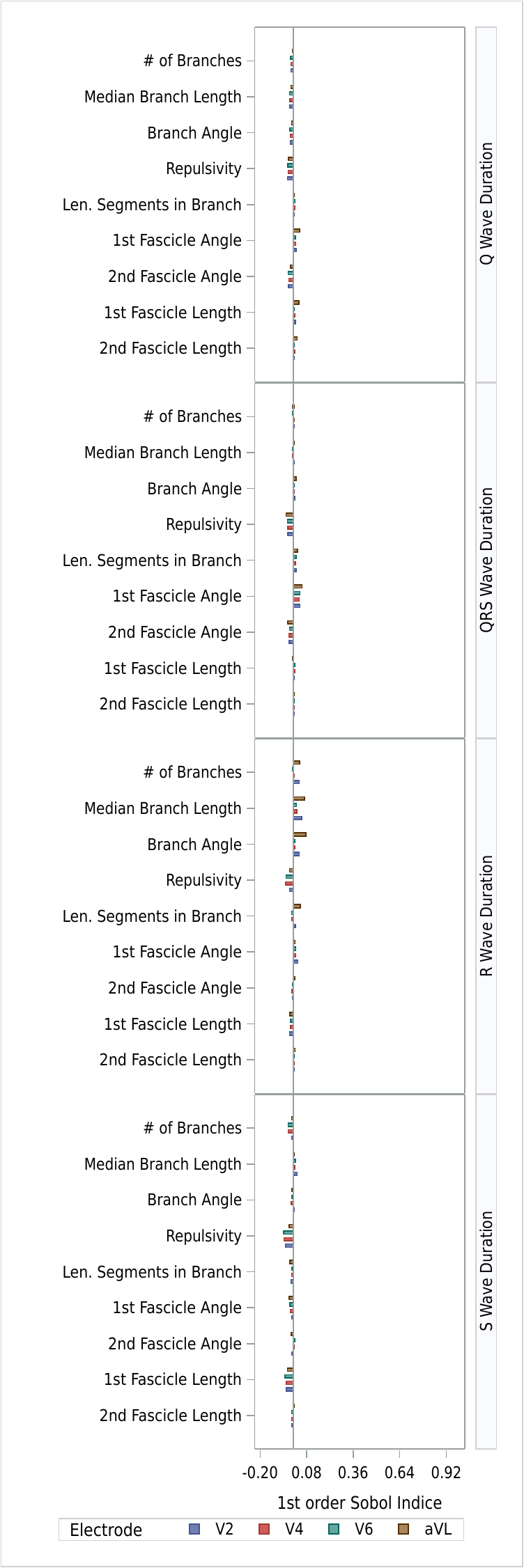}%
    \includegraphics[width=0.33\textwidth]{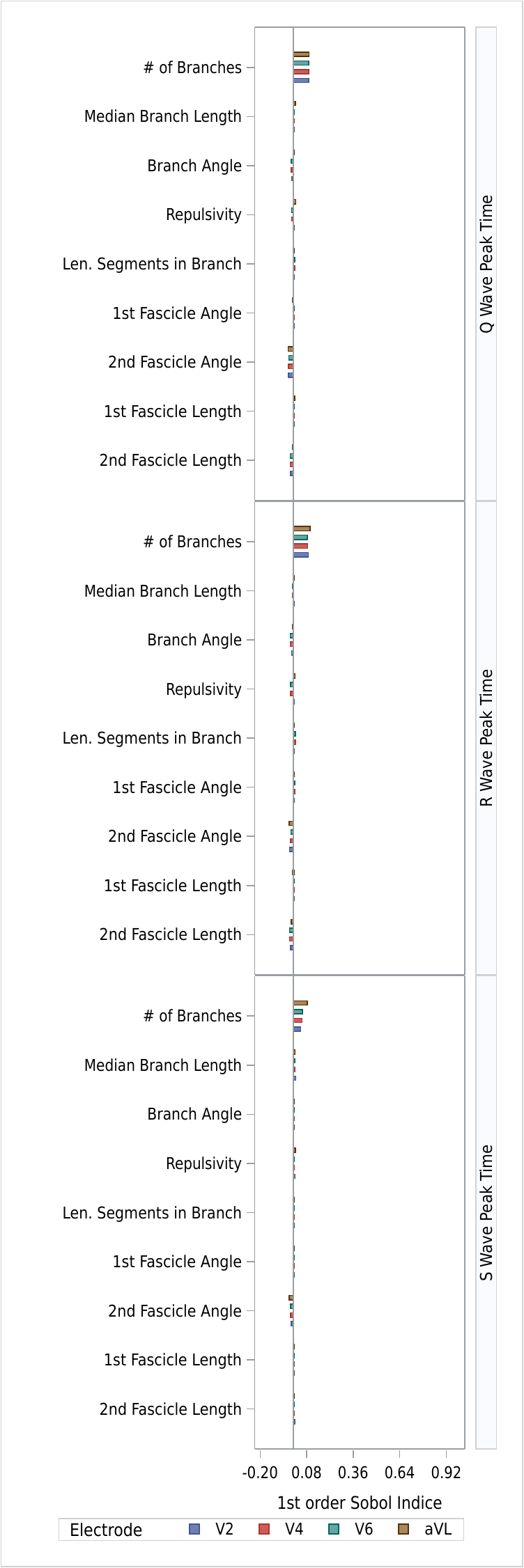}%
    \includegraphics[width=0.33\textwidth]{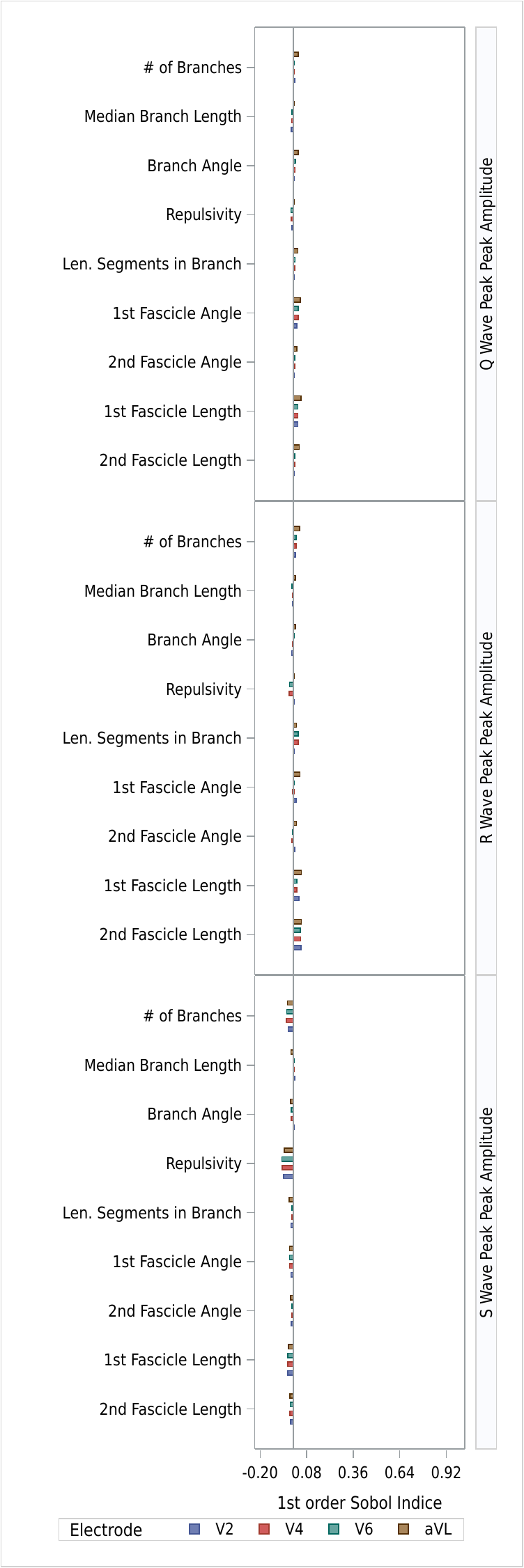}
    \caption{First order Sobol sensitivity indices for the ten quantities of interest across all HPS parameters}
    \label{fig:first_order_sobol}
\end{figure}

\noindent When analyzing the Sobol sensitivity indices to assess the influence of model parameters on these QOIs, very few significant first-order (main) effects were identified. This suggests that individual parameter variations do not substantially contribute to the limited variability observed in the QOIs. All first order indices are plotted in Figure \ref{fig:first_order_sobol}.\\

\noindent For peak amplitudes, no significant first-order effects were found. The largest index was observed for the R-wave amplitude in lead aVL, where the second fascicle length accounted for approximately 4.7\% of the observed variability (\(S_1 = 0.0474\); see Table~\ref{tab:first_order_sobol_ampl} in the Supplementary Tables). In the case of wave durations, branch angle had the largest influence on R-wave duration in lead aVL, accounting approximately 7.47\% of the variability in this distribution. Additionally, significant first-order effects were observed for median branch length on R-wave duration in leads V2 (\(S_1 = 0.503\)) and V6 (\(S_1 = 0.675\)), and for the first fascicle angle on overall QRS duration in lead aVL (\(S_1 = 0.0528\)). However, as shown in Figure~\ref{fig:sig_wave_dur_sob}, most of these significant \(S_1\) values only marginally exceed the 0.05 threshold and are not consistently observed across all four leads. Table~\ref{tab:first_order_wave_durations} in the Supplementary Tables provides all first-order indices for these parameters and QOIs across all leads.\\

\begin{figure}[h!]
    \centering
    \includegraphics[width=0.33\textwidth]{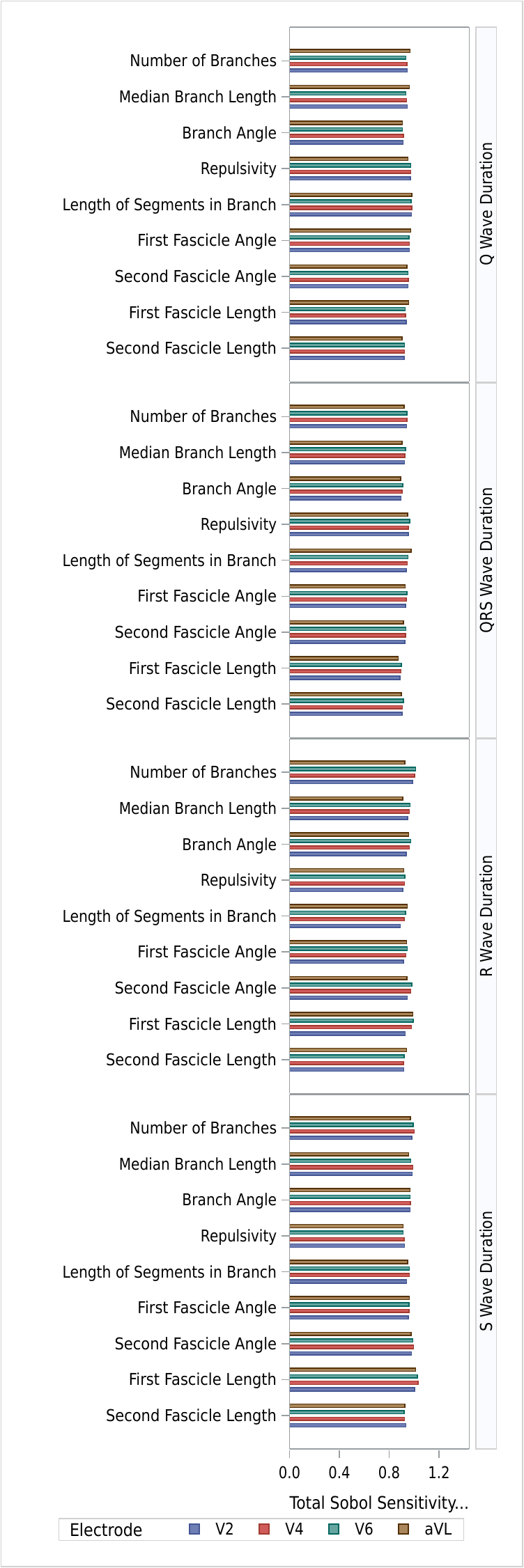}%
    \includegraphics[width=0.33\textwidth]{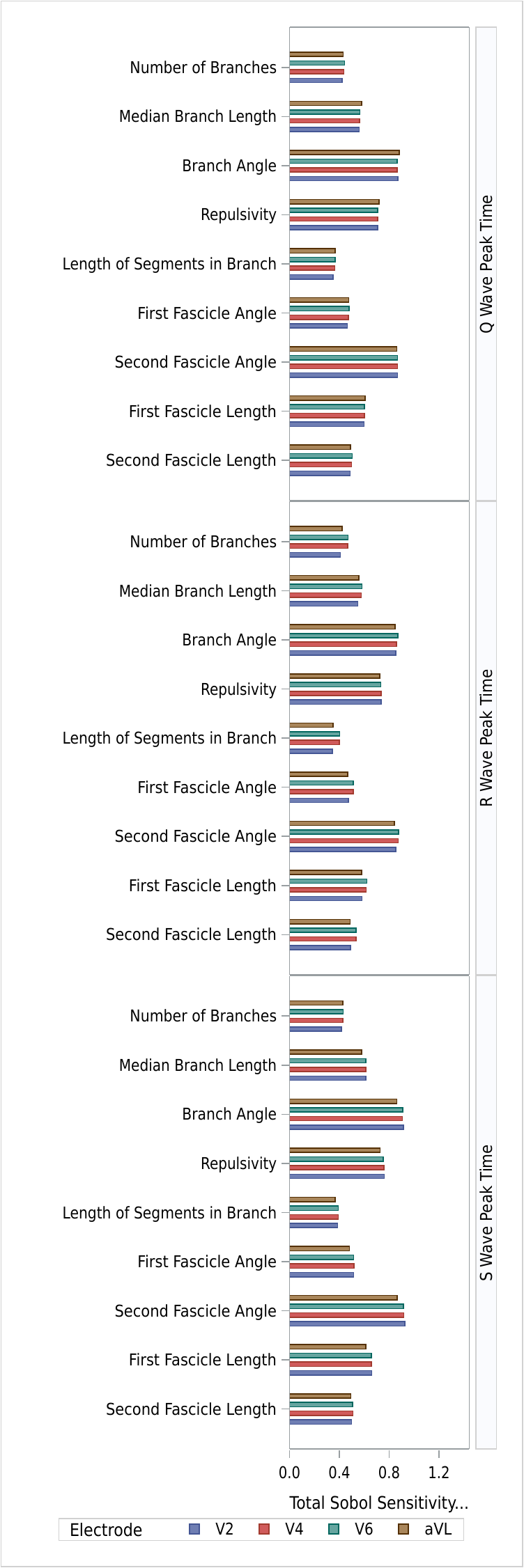}%
    \includegraphics[width=0.33\textwidth]{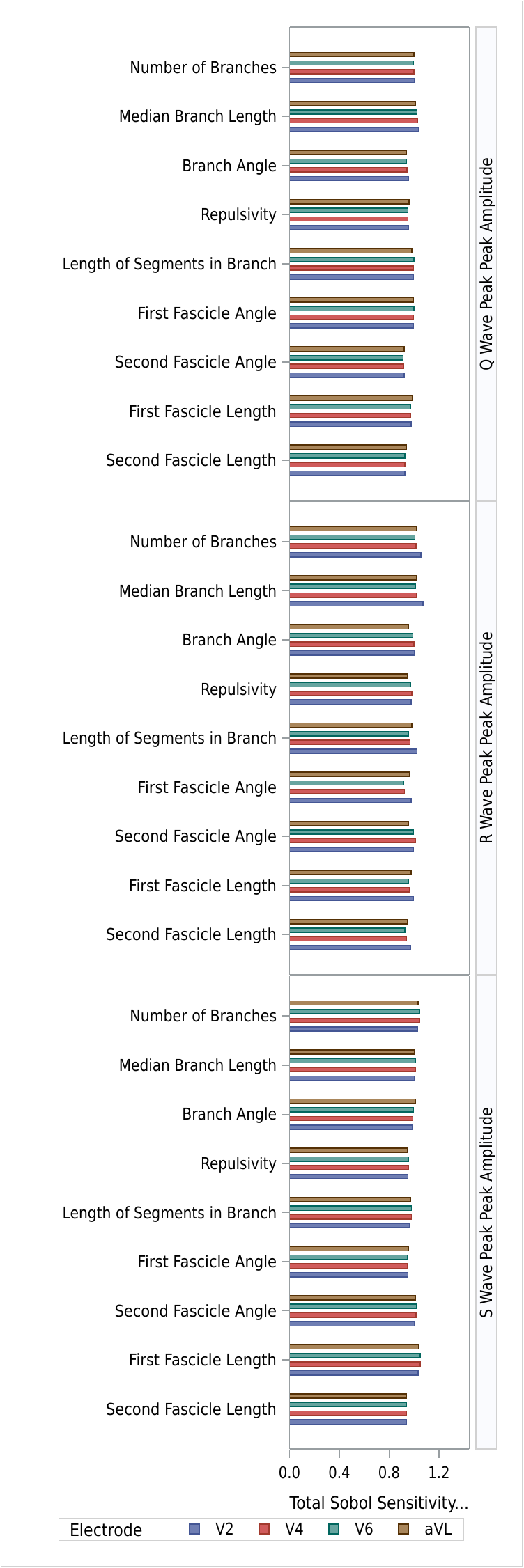}
    \caption{Total Sobol sensitivity indices for the ten quantities of interest across all HPS parameters.}
    \label{fig:total_sobol_indices_bar} 
\end{figure}

\noindent While the main effects are generally small and statistically insignificant, the total-order Sobol indices are substantially larger. A chart of all total Sobol sensitivity indices is presented in Figure~\ref{fig:total_sobol_indices_bar}. For these sets of quantities of interest, all $S_\text{T}$ values either equaled one or contained one in their confidence intervals. This indicates that the observed variability in these QOIs is primarily driven by higher-order interaction effects among the HPS parameters, rather than by individual parameters alone.\\

\begin{figure}[h!]
    \centering
    \includegraphics[width=0.5\textwidth]{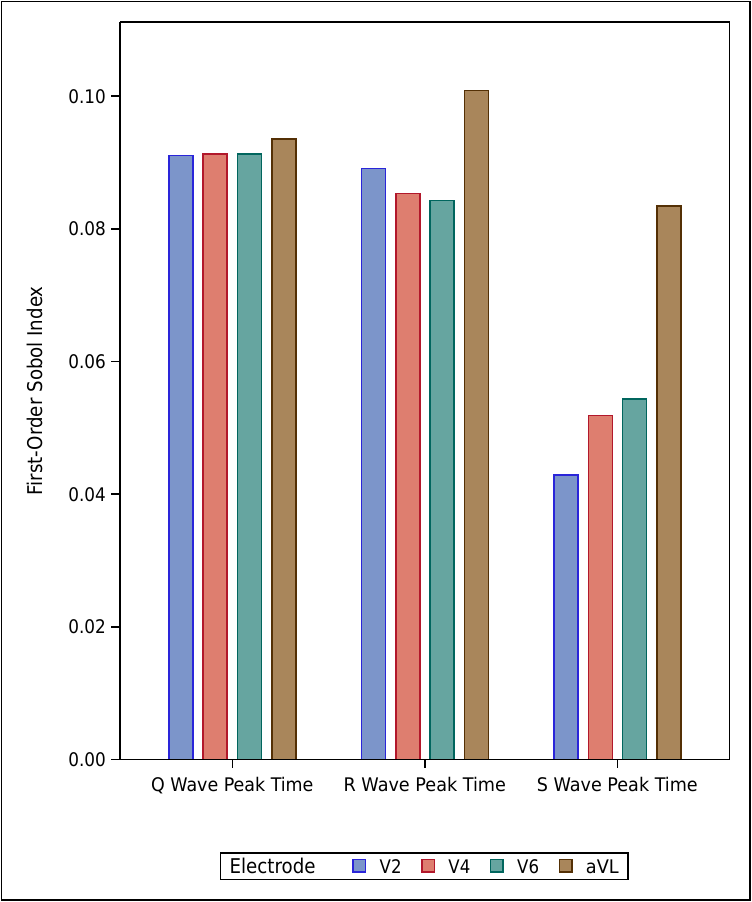}%
    \includegraphics[width=0.5\textwidth]{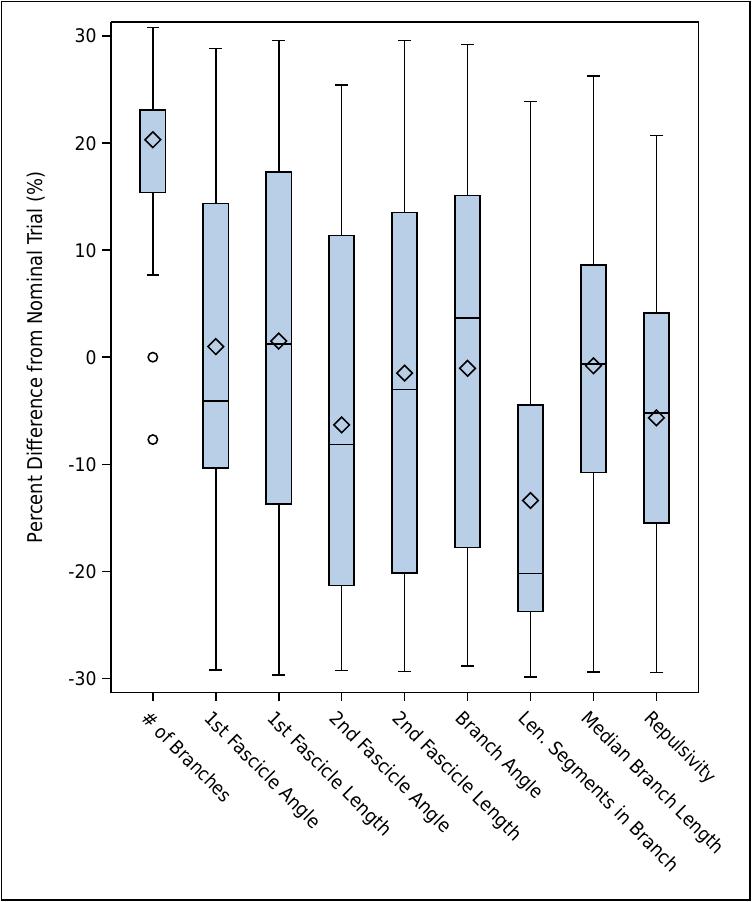}
    \caption{\textit{(Left)} First‐order Sobol sensitivity indices for the number of branches across all peak‐time simulations. \textit{(Right)} Percent differences from nominal values for each of the 9 parameters across the 25 outlier trials, with the y-axis indicating the percent deviation from nominal values in each simulation.
  }
    \label{fig:num_branches_sobol} 
\end{figure}

\noindent Unlike the $S_1$ values for wave duration and peak amplitude distributions, peak times exhibited moderately large first-order indices that were consistent across all lead fields. As shown in Figure~\ref{fig:num_branches_sobol}, the number of branches had a significant and moderately strong effect across all leads and individual peak times, except for the S-wave peak time in lead V2 ($S_1 = 0.0429$). The largest effect was observed in the R-wave of lead aVL ($S_1 = 0.101$), where the number of branches accounted for approximately 10.1\% of the variability in peak times. Table~\ref{tab:first_order_sobol_times} presents all first-order index values for peak times. \\

\noindent When analyzing the distributions of parameters in the peak times outlier trials, there were more outliers associated with a greater number of branches, which supports the main effect observed in the Sobol analysis and suggests that the number of branches in the network plays a role in generating these outliers. However, while outliers tended to have a greater number of branches, there were also many non-outlier trials with the same number of branches, indicating that the number of branches alone cannot fully explain the occurrence of early peak times.\\

\begin{figure}[h!]
    \centering
    \includegraphics[width=1.0\linewidth]{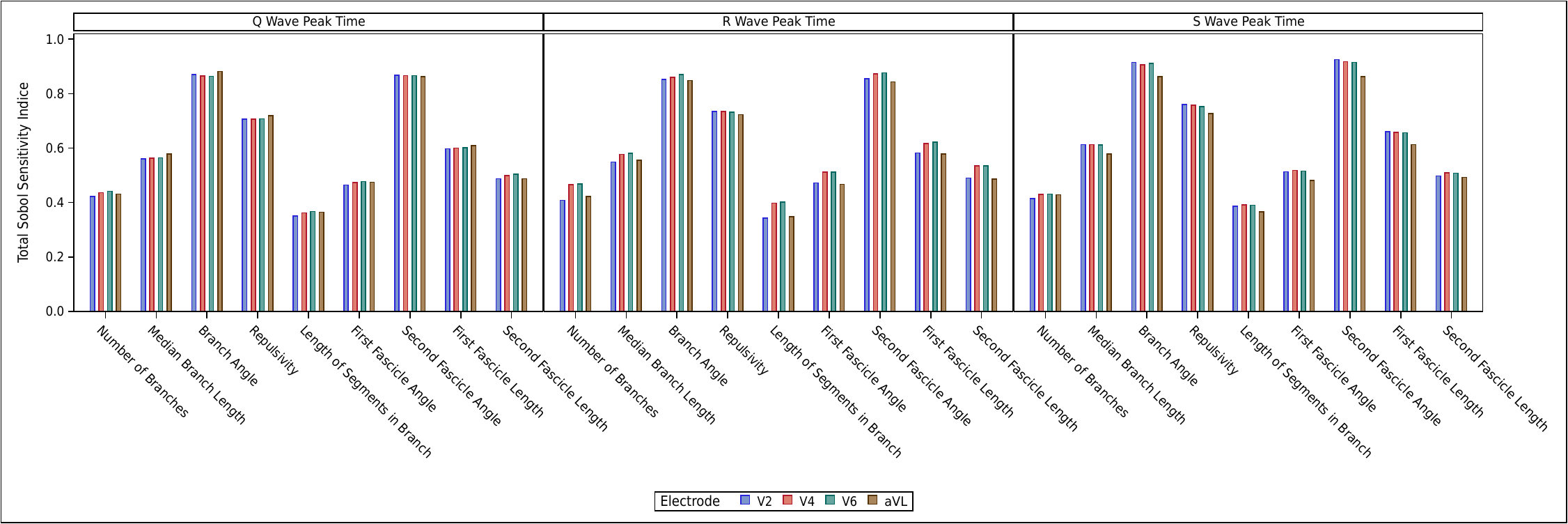}
    \caption{Total Sobol Sensitivity Indices for all Parameters across Peak Times}
    \label{fig:Total_Sobol_Times}
\end{figure}

\noindent Further differentiating peak times from the other QOI, there is notable heterogeneity among \(S_\text{T}\) values for peak times, as seen in Figure~\ref{fig:Total_Sobol_Times}. While all \(S_\text{T}\) values are significant, those for branch angle and second fascicle angle were the highest, with nearly all falling in the range of 0.85–0.95, indicating that these terms are highly involved in interactions that drive variability. Repulsivity, although not as large, also exhibited a very high \(S_\text{T}\) values, with all falling in the range of 0.7-0.8, suggesting that it is also highly involved in interactions contributing to variability in peak times. 

\section{Discussion}

\noindent Nearly all trials of peak amplitudes and wave durations exhibited minimal variation, suggesting that these aspects of QRS morphology are largely robust to minor variations in HPS structure. Several metrics of variability showed very low variability due to changes in the HPS. Excluding outliers, almost all data points fell within 2.5 standard deviations of the mean, with the majority concentrated within 0.5–0.75 standard deviations. The greatest standard deviations for peak amplitudes and wave durations were 0.183 mV and 3.88 ms, respectively, and $\text{CV}$ values remained low. This low level of variability from the means for these QOI remained within the expected range for a normal QRS complex, indicating that the observed variation is not clinically significant for distinguishing abnormal QRS complexes from normal ones in disease diagnosis \cite{normal_ecg_and_disease_specs}.\\

\noindent Although most trials exhibited low variability, there were a few extreme outliers that inflated the overall spread, as reflected in the $z_{\text{range}}$ values. A single trial was identified as a significant outlier across multiple distributions for these QOIs and is visualized in Figure \ref{fig:outlier_trial} in the Supplementary Figures. The QRS complex of this trial exhibited a markedly large S-wave and an abnormally small R-wave, particularly in leads V2, V4, and aVL. This pattern is notable, as a deeper and longer S-wave is a key feature in diagnosing cardiac conditions such as left ventricular hypertrophy and right bundle branch block \cite{LVH_diagnosis_criteria, normal_ecg_and_disease_specs}.\\

\noindent Sobol analysis revealed consistently low main effects across all parameters for these QOIs. Four parameters exhibited marginal significance for wave durations, however, their effects were inconsistent across the four leads. In contrast, \( S_\text{T} \) values were approximately one for all parameters, indicating that the variability in these QOIs is primarily driven by interactions among HPS parameters rather than individual parameter effects. Furthermore, this suggests that the observed outliers in these distributions, such as the one previously discussed, result from specific parameter combinations that caused interaction effects rather than isolated parameter changes. Thus, while minor variations in individual parameters have a limited impact, specific combinations of parameter values can cause interaction effects that lead to clinically significant variations in peak amplitude and wave durations. \\

\noindent Unlike wave durations and peak amplitudes, peak times exhibited a distinctive pattern in outlier distribution. While overall variability excluding outliers remained low and comparable to other QOIs, the presence of outliers clustering around a secondary central point—rather than near the bounds of the primary distribution—significantly inflated the overall spread. Visualization of these trials revealed that, while the QRS complexes were morphologically similar to those in the primary distribution, they formed significantly earlier. Although P-wave characterization is beyond the scope of this study, the early formation of these QRS complexes appeared to obscure P-wave formation, effectively masking the corresponding ECG waveforms (see Figure~\ref{fig:outlier_trial}). This premature formation of the QRS complexes can hinder accurate modeling of cardiac arrhythmias, particularly conditions such as atrial fibrillation, which rely on the presence or absence of P-waves as a diagnostic criterion \cite{afib_diagnoses}. HPS configurations that produce overly rapid QRS formation may lead to inaccurate ECG representations, potentially affecting the diagnostic utility of CDTs for detecting these arrhythmias. \\

\noindent Sobol analysis revealed a significant and moderately large main effects for the number of branches in the HPS across all leads and almost all peak times. This suggests that variation in the number of branches has a meaningful impact on the timing of QRS complex formation. Additionally, unlike wave durations and peak amplitudes, \(S_\text{T}\) indices exhibited significant heterogeneity. Rather than variability being entirely caused by high-level interactions involving all HPS parameters, branch angle, second fascicle angle, and repulsivity were substantially more influential than other parameters, indicating that these HPS parameters are more involved in interactions contributing to the variability observed in peak timings. While the remaining parameters did have significant and large \(S_\text{T}\), indicating they strongly impact variability due to interactions, they were not as highly involved as these three. Thus, peak timing variability is largely driven by both the direct effects of variation in the number of branches and interaction effects between parameters, with branch angle, repulsivity, and fascicle angle being the most involved in these interactions.

\subsection{Significance and Future Work}

\noindent The variability of HPS structure in human populations remains an open question in cardiology. The true distributions of key structural features in actual individuals are largely unknown, posing a challenge for developing accurate cardiac models. Prior studies have emphasized the importance of assessing how variations in HPS structure influence model derived outputs like the ECG, yet a comprehensive quantification of this variability has been lacking \cite{including_HPS_CDT, second_sigma_b}. This study addresses this gap by systematically quantifying the variability introduced by HPS structural parameters and identifying the key drivers of this variability. By doing so, our findings represent a crucial step toward developing cardiac models that are robust to underlying anatomical uncertainties, ultimately enhancing their reliability for both research and clinical applications.\\

\noindent Furthermore, this study represents one of the first comprehensive global sensitivity analyses conducted on a complete 3D cardiac electrophysiology model, enabling direct comparison of observed variability to clinical settings. While computational cost was not the primary focus of this work, Sobol sensitivity analysis is inherently computationally intensive and few studies have applied it at the scale of a full 3D heart model, highlighting the novelty of our approach \cite{Pras_GSA_1}. By leveraging this rigorous sensitivity analysis, our study provides deeper insights into the impact of structural variability on cardiac model predictions, paving the way for more robust and physiologically relevant simulations.\\

\noindent While this study provides valuable insights into the impact of variability in HPS structure on cardiac model outputs, further research is needed to fully characterize the factors influencing modeled cardiac electrophysiology. Specifically, although significant main effects and interactions among structural HPS parameters were identified, the influence of electrophysiological parameters and their interactions with structural parameters was not explored in this study. Additionally, while this work focuses on cardiac electrophysiology, further investigation is needed into the role of HPSs in coupled electromechanics, including their influence on local tissue strain distributions and global pressure-volume loops \cite{fractal_tree_citation}. Quantifying these effects will build on the current study, improving our understanding of model robustness and advancing the development of more physiologically accurate cardiovascular digital twin models. \\
\section{Conclusion}
\noindent Diagnostically relevant QRS complex features exhibit overall low sensitivity to minor variations in individual HPS structural parameters. Small perturbations in individual structural parameters did not lead to clinically significant changes in wave durations or amplitudes;  however, their combined effects caused pronounced morphological alterations with clinical implications, as observed in outlier trials.  Similarly, peak timings is also largely insensitive to minor perturbations in HPS structural parameters, however our results indicate that some variation in HPS structure can cause premature QRS complex formation. This premature QRS formation has substantial downstream effect, particularly in obscuring other ECG waveforms, such as the P-wave. We found that individual variation in the number of branches in the network, alongside interactions between HPS parameters drive this observed variability. Interactions affecting variability in peak timing were primarily centered around branch and fascicle angles, and to a lesser extent branch repulsivity. However, other parameters also had significant interaction effects. While future models should account for these potential sources of variability introduced by the HPS structure, this study indicates that minor structural differences between a healthy patient’s HPS and that of a generic model are unlikely to significantly impact model fidelity or clinical interpretation when both systems are physiologically normal.

\section{Acknowledgments}

\noindent Computations were performed using facilities provided by the University of North Carolina at Chapel Hill through the Research Computing division of UNC Information Technology Services. During the preparation of this work, the authors used ChatGPT to improve language and readability. After using this tool, the authors reviewed and edited the content as needed and take full responsibility for the content of this publication.

\section{Funding}

\noindent This work was supported in part by the National Institutes of Health under Grant R01HL157631 and U01HL143336, and in part by the National Science Foundation under Grant OAC 1931516.

\renewcommand\refname{REFERENCES}
\bibliographystyle{elsarticle-num-names}
\bibliography{main}            

\begin{thebibliography}{2}
\expandafter\ifx\csname natexlab\endcsname\relax\def\natexlab#1{#1}\fi
\providecommand{\url}[1]{\texttt{#1}}
\providecommand{\href}[2]{#2}
\providecommand{\path}[1]{#1}
\providecommand{\DOIprefix}{doi:}
\providecommand{\ArXivprefix}{arXiv:}
\providecommand{\URLprefix}{URL: }
\providecommand{\Pubmedprefix}{pmid:}
\providecommand{\doi}[1]{\href{http://dx.doi.org/#1}{\path{#1}}}
\providecommand{\Pubmed}[1]{\href{pmid:#1}{\path{#1}}}
\providecommand{\bibinfo}[2]{#2}
\ifx\xfnm\relax \def\xfnm[#1]{\unskip,\space#1}\fi
\bibitem[{Bishop and Plank(2011)}]{ecg_eq_1}
\bibinfo{author}{M.~J. Bishop}, \bibinfo{author}{G.~Plank},
\newblock \bibinfo{title}{Bidomain ecg simulations using an augmented
  monodomain model for the cardiac source},
\newblock \bibinfo{journal}{IEEE Transactions on Biomedical Engineering}
  \bibinfo{volume}{58} (\bibinfo{year}{2011}) \bibinfo{pages}{2297--2307}.
  \DOIprefix\doi{10.1109/TBME.2011.2148718}.
\bibitem[{Ten~Tusscher and Panfilov(2006)}]{TP06_model}
\bibinfo{author}{K.~Ten~Tusscher}, \bibinfo{author}{A.~Panfilov},
\newblock \bibinfo{title}{Alternans and spiral breakup in a human ventricular
  tissue model},
\newblock \bibinfo{journal}{Am J Physiol Heart Circ Physiol}
  \bibinfo{volume}{291} (\bibinfo{year}{2006}) \bibinfo{pages}{H1088--H1100}.
  \DOIprefix\doi{10.1152/ajpheart.00109.2006}.

\end{thebibliography}


\begin{thebibliography}{47}
\expandafter\ifx\csname natexlab\endcsname\relax\def\natexlab#1{#1}\fi
\providecommand{\url}[1]{\texttt{#1}}
\providecommand{\href}[2]{#2}
\providecommand{\path}[1]{#1}
\providecommand{\DOIprefix}{doi:}
\providecommand{\ArXivprefix}{arXiv:}
\providecommand{\URLprefix}{URL: }
\providecommand{\Pubmedprefix}{pmid:}
\providecommand{\doi}[1]{\href{http://dx.doi.org/#1}{\path{#1}}}
\providecommand{\Pubmed}[1]{\href{pmid:#1}{\path{#1}}}
\providecommand{\bibinfo}[2]{#2}
\ifx\xfnm\relax \def\xfnm[#1]{\unskip,\space#1}\fi
\bibitem[{Martin et~al.(2024)Martin, Aday, Almarzooq, Anderson, Arora, Avery,
  Baker-Smith, Gibbs, Beaton, Boehme, Commodore-Mensah, Currie, Elkind,
  Evenson, Generoso, Heard, Hiremath, Johansen, Kalani, Kazi, Ko, Liu, Magnani,
  Michos, Mussolino, Navaneethan, Parikh, Perman, Poudel, Rezk-Hanna, Roth,
  Shah, St-Onge, Thacker, Tsao, Urbut, Spall, Voeks, Wang, Wong, Wong, Yaffe,
  Palaniappan, on~behalf of the American Heart Association Council~on
  Epidemiology, Committee, and Subcommittee}]{heart_disease_leading_death}
\bibinfo{author}{S.~S. Martin}, \bibinfo{author}{A.~W. Aday},
  \bibinfo{author}{Z.~I. Almarzooq}, \bibinfo{author}{C.~A. Anderson},
  \bibinfo{author}{P.~Arora}, \bibinfo{author}{C.~L. Avery},
  \bibinfo{author}{C.~M. Baker-Smith}, \bibinfo{author}{B.~B. Gibbs},
  \bibinfo{author}{A.~Z. Beaton}, \bibinfo{author}{A.~K. Boehme},
  \bibinfo{author}{Y.~Commodore-Mensah}, \bibinfo{author}{M.~E. Currie},
  \bibinfo{author}{M.~S. Elkind}, \bibinfo{author}{K.~R. Evenson},
  \bibinfo{author}{G.~Generoso}, \bibinfo{author}{D.~G. Heard},
  \bibinfo{author}{S.~Hiremath}, \bibinfo{author}{M.~C. Johansen},
  \bibinfo{author}{R.~Kalani}, \bibinfo{author}{D.~S. Kazi},
  \bibinfo{author}{D.~Ko}, \bibinfo{author}{J.~Liu}, \bibinfo{author}{J.~W.
  Magnani}, \bibinfo{author}{E.~D. Michos}, \bibinfo{author}{M.~E. Mussolino},
  \bibinfo{author}{S.~D. Navaneethan}, \bibinfo{author}{N.~I. Parikh},
  \bibinfo{author}{S.~M. Perman}, \bibinfo{author}{R.~Poudel},
  \bibinfo{author}{M.~Rezk-Hanna}, \bibinfo{author}{G.~A. Roth},
  \bibinfo{author}{N.~S. Shah}, \bibinfo{author}{M.-P. St-Onge},
  \bibinfo{author}{E.~L. Thacker}, \bibinfo{author}{C.~W. Tsao},
  \bibinfo{author}{S.~M. Urbut}, \bibinfo{author}{H.~G.~V. Spall},
  \bibinfo{author}{J.~H. Voeks}, \bibinfo{author}{N.-Y. Wang},
  \bibinfo{author}{N.~D. Wong}, \bibinfo{author}{S.~S. Wong},
  \bibinfo{author}{K.~Yaffe}, \bibinfo{author}{L.~P. Palaniappan},
  \bibinfo{author}{on~behalf of the American Heart Association Council~on
  Epidemiology}, \bibinfo{author}{P.~S. Committee}, \bibinfo{author}{S.~S.
  Subcommittee},
\newblock \bibinfo{title}{2024 heart disease and stroke statistics: A report of
  us and global data from the american heart association},
\newblock \bibinfo{journal}{Circulation} \bibinfo{volume}{149}
  (\bibinfo{year}{2024}) \bibinfo{pages}{e347--e913}.
  \DOIprefix\doi{10.1161/CIR.0000000000001209}.
\bibitem[{John et~al.(2012)John, Tedrow, Koplan, Albert, Epstein, Sweeney,
  Miller, Michaud, and Stevenson}]{arrythmias_death}
\bibinfo{author}{R.~M. John}, \bibinfo{author}{U.~B. Tedrow},
  \bibinfo{author}{B.~A. Koplan}, \bibinfo{author}{C.~M. Albert},
  \bibinfo{author}{L.~M. Epstein}, \bibinfo{author}{M.~O. Sweeney},
  \bibinfo{author}{A.~L. Miller}, \bibinfo{author}{G.~F. Michaud},
  \bibinfo{author}{W.~G. Stevenson},
\newblock \bibinfo{title}{Ventricular arrhythmias and sudden cardiac death},
\newblock \bibinfo{journal}{The Lancet} \bibinfo{volume}{380}
  (\bibinfo{year}{2012}) \bibinfo{pages}{1520--1529}.
  \DOIprefix\doi{10.1016/S0140-6736(12)61413-5}.
\bibitem[{Corral-Acero et~al.(2020)Corral-Acero, Margara, Marciniak, Rodero,
  Loncaric, Feng, Gilbert, Fernandes, Bukhari, Wajdan, Martinez, Santos,
  Shamohammdi, Luo, Westphal, Leeson, DiAchille, Gurev, Mayr, Geris,
  Pathmanathan, Morrison, Cornelussen, Prinzen, Delhaas, Doltra, Sitges,
  Vigmond, Zacur, Grau, Rodriguez, Remme, Niederer, Mortier, McLeod, Potse,
  Pueyo, Bueno-Orovio, and Lamata}]{digital_twin_definition}
\bibinfo{author}{J.~Corral-Acero}, \bibinfo{author}{F.~Margara},
  \bibinfo{author}{M.~Marciniak}, \bibinfo{author}{C.~Rodero},
  \bibinfo{author}{F.~Loncaric}, \bibinfo{author}{Y.~Feng},
  \bibinfo{author}{A.~Gilbert}, \bibinfo{author}{J.~F. Fernandes},
  \bibinfo{author}{H.~A. Bukhari}, \bibinfo{author}{A.~Wajdan},
  \bibinfo{author}{M.~V. Martinez}, \bibinfo{author}{M.~S. Santos},
  \bibinfo{author}{M.~Shamohammdi}, \bibinfo{author}{H.~Luo},
  \bibinfo{author}{P.~Westphal}, \bibinfo{author}{P.~Leeson},
  \bibinfo{author}{P.~DiAchille}, \bibinfo{author}{V.~Gurev},
  \bibinfo{author}{M.~Mayr}, \bibinfo{author}{L.~Geris},
  \bibinfo{author}{P.~Pathmanathan}, \bibinfo{author}{T.~Morrison},
  \bibinfo{author}{R.~Cornelussen}, \bibinfo{author}{F.~Prinzen},
  \bibinfo{author}{T.~Delhaas}, \bibinfo{author}{A.~Doltra},
  \bibinfo{author}{M.~Sitges}, \bibinfo{author}{E.~J. Vigmond},
  \bibinfo{author}{E.~Zacur}, \bibinfo{author}{V.~Grau},
  \bibinfo{author}{B.~Rodriguez}, \bibinfo{author}{E.~W. Remme},
  \bibinfo{author}{S.~Niederer}, \bibinfo{author}{P.~Mortier},
  \bibinfo{author}{K.~McLeod}, \bibinfo{author}{M.~Potse},
  \bibinfo{author}{E.~Pueyo}, \bibinfo{author}{A.~Bueno-Orovio},
  \bibinfo{author}{P.~Lamata},
\newblock \bibinfo{title}{The ‘digital twin’ to enable the vision of
  precision cardiology},
\newblock \bibinfo{journal}{European Heart Journal} \bibinfo{volume}{41}
  (\bibinfo{year}{2020}) \bibinfo{pages}{4556--4564}.
  \DOIprefix\doi{10.1093/eurheartj/ehaa159}.
\bibitem[{Viceconti et~al.(2016)Viceconti, Marco, Henney, Adrian,
  Morley-Fletcher, and Edward}]{device_testing}
\bibinfo{author}{Viceconti}, \bibinfo{author}{Marco}, \bibinfo{author}{Henney},
  \bibinfo{author}{Adrian}, \bibinfo{author}{Morley-Fletcher},
  \bibinfo{author}{Edward},
\newblock \bibinfo{title}{In silico clinical trials: how computer simulation
  will transform the biomedical industry},
\newblock \bibinfo{journal}{International Journal of Clinical Trials}
  \bibinfo{volume}{3} (\bibinfo{year}{2016}) \bibinfo{pages}{37--46}.
  \DOIprefix\doi{10.18203/2349-3259.ijct20161408}.
\bibitem[{Gillette et~al.(2021)Gillette, Gsell, Prassl, Karabelas, Reiter,
  Reiter, Grandits, Payer, Štern, Urschler, Bayer, Augustin, Neic, Pock,
  Vigmond, and Plank}]{first_sigma_b}
\bibinfo{author}{K.~Gillette}, \bibinfo{author}{M.~A. Gsell},
  \bibinfo{author}{A.~J. Prassl}, \bibinfo{author}{E.~Karabelas},
  \bibinfo{author}{U.~Reiter}, \bibinfo{author}{G.~Reiter},
  \bibinfo{author}{T.~Grandits}, \bibinfo{author}{C.~Payer},
  \bibinfo{author}{D.~Štern}, \bibinfo{author}{M.~Urschler},
  \bibinfo{author}{J.~D. Bayer}, \bibinfo{author}{C.~M. Augustin},
  \bibinfo{author}{A.~Neic}, \bibinfo{author}{T.~Pock}, \bibinfo{author}{E.~J.
  Vigmond}, \bibinfo{author}{G.~Plank},
\newblock \bibinfo{title}{A framework for the generation of digital twins of
  cardiac electrophysiology from clinical 12-leads ecgs},
\newblock \bibinfo{journal}{Medical Image Analysis} \bibinfo{volume}{71}
  (\bibinfo{year}{2021}) \bibinfo{pages}{102080}.
  \DOIprefix\doi{https://doi.org/10.1016/j.media.2021.102080}.
\bibitem[{Grandits et~al.(2025)Grandits, Gillette, Plank, and
  Pezzuto}]{ecg_digital_twin}
\bibinfo{author}{T.~Grandits}, \bibinfo{author}{K.~Gillette},
  \bibinfo{author}{G.~Plank}, \bibinfo{author}{S.~Pezzuto},
  \bibinfo{title}{Accurate and efficient cardiac digital twin from surface
  ecgs: Insights into identifiability of ventricular conduction system},
  \bibinfo{year}{2025}. \URLprefix \url{https://arxiv.org/abs/2411.00165}.
  \href{http://arxiv.org/abs/2411.00165}{{\tt arXiv:2411.00165}}.
\bibitem[{Viola et~al.(2023)Viola, Del~Corso, De~Paulis, and
  Verzicco}]{ecg_digital_twin_2}
\bibinfo{author}{F.~Viola}, \bibinfo{author}{G.~Del~Corso},
  \bibinfo{author}{R.~De~Paulis}, \bibinfo{author}{R.~Verzicco},
\newblock \bibinfo{title}{Gpu accelerated digital twins of the human heart open
  new routes for cardiovascular research},
\newblock \bibinfo{journal}{Scientific Reports} \bibinfo{volume}{13}
  (\bibinfo{year}{2023}) \bibinfo{pages}{8230}.
  \DOIprefix\doi{10.1038/s41598-023-34098-8}.
\bibitem[{Boyden(2018)}]{what_are_purkinje}
\bibinfo{author}{P.~A. Boyden},
\newblock \bibinfo{title}{Purkinje physiology and pathophysiology},
\newblock \bibinfo{journal}{Journal of Interventional Cardiac
  Electrophysiology} \bibinfo{volume}{52} (\bibinfo{year}{2018})
  \bibinfo{pages}{255--262}. \DOIprefix\doi{10.1007/s10840-018-0414-3},
  \bibinfo{note}{epub 2018 Jul 28}.
\bibitem[{Ashley and Niebauer(2004)}]{role_of_HPS}
\bibinfo{author}{E.~A. Ashley}, \bibinfo{author}{J.~Niebauer},
  \bibinfo{title}{Conquering the ECG}, \bibinfo{publisher}{Remedica},
  \bibinfo{address}{London}, \bibinfo{year}{2004}. \URLprefix
  \url{https://www.ncbi.nlm.nih.gov/books/NBK2214/}, \bibinfo{note}{available
  from: \url{https://www.ncbi.nlm.nih.gov/books/NBK2214/}}.
\bibitem[{Sattar and Chhabra(2023)}]{role_of_HPS_2}
\bibinfo{author}{Y.~Sattar}, \bibinfo{author}{L.~Chhabra},
\newblock \bibinfo{title}{Electrocardiogram},
\newblock in: \bibinfo{booktitle}{StatPearls [Internet]},
  \bibinfo{publisher}{StatPearls Publishing}, \bibinfo{address}{Treasure Island
  (FL)}, \bibinfo{year}{2023}. \URLprefix
  \url{https://www.ncbi.nlm.nih.gov/books/NBK549803/}, \bibinfo{note}{updated
  2023 Jun 5}.
\bibitem[{Dubin(1996)}]{purkinje_contraction_sync}
\bibinfo{author}{D.~Dubin}, \bibinfo{title}{Rapid Interpretation of EKG's},
  \bibinfo{publisher}{Cover Publishing Company}, \bibinfo{year}{1996}.
\bibitem[{Itoh and Yamada(2018)}]{HPS_QRS_disease}
\bibinfo{author}{T.~Itoh}, \bibinfo{author}{T.~Yamada},
\newblock \bibinfo{title}{Multifocal ventricular arrhythmias originating from
  the his-purkinje system},
\newblock \bibinfo{journal}{JACC: Clinical Electrophysiology}
  \bibinfo{volume}{4} (\bibinfo{year}{2018}) \bibinfo{pages}{1248--1260}.
  \DOIprefix\doi{10.1016/j.jacep.2018.06.015}.
\bibitem[{McAnulty and Rahimtoola(1984)}]{Diagnosing_BBB}
\bibinfo{author}{J.~H. McAnulty}, \bibinfo{author}{S.~H. Rahimtoola},
\newblock \bibinfo{title}{Bundle branch block},
\newblock \bibinfo{journal}{Progress in Cardiovascular Diseases}
  \bibinfo{volume}{26} (\bibinfo{year}{1984}) \bibinfo{pages}{333--354}.
  \DOIprefix\doi{https://doi.org/10.1016/0033-0620(84)90009-4}.
\bibitem[{Surawicz et~al.(2009)Surawicz, Childers, Deal, and
  Gettes}]{normal_ecg_and_disease_specs}
\bibinfo{author}{B.~Surawicz}, \bibinfo{author}{R.~Childers},
  \bibinfo{author}{B.~J. Deal}, \bibinfo{author}{L.~S. Gettes},
\newblock \bibinfo{title}{Aha/accf/hrs recommendations for the standardization
  and interpretation of the electrocardiogram},
\newblock \bibinfo{journal}{Circulation} \bibinfo{volume}{119}
  (\bibinfo{year}{2009}) \bibinfo{pages}{e235--e240}.
  \DOIprefix\doi{10.1161/CIRCULATIONAHA.108.191095}.
\bibitem[{Jaffery et~al.(2024)Jaffery, Melki, Slabaugh, Good, and
  Roney}]{scans_for_geometries}
\bibinfo{author}{O.~A. Jaffery}, \bibinfo{author}{L.~Melki},
  \bibinfo{author}{G.~Slabaugh}, \bibinfo{author}{W.~W. Good},
  \bibinfo{author}{C.~H. Roney},
\newblock \bibinfo{title}{A review of personalised cardiac computational
  modelling using electroanatomical mapping data},
\newblock \bibinfo{journal}{Arrhythmia \& Electrophysiology Review
  2024;13:e08.}  (\bibinfo{year}{2024}). \DOIprefix\doi{10.15420/aer.2023.25}.
\bibitem[{Çetingül et~al.(2011)Çetingül, Plank, Trayanova, and
  Vidal}]{cant_see_HPS}
\bibinfo{author}{H.~E. Çetingül}, \bibinfo{author}{G.~Plank},
  \bibinfo{author}{N.~A. Trayanova}, \bibinfo{author}{R.~Vidal},
\newblock \bibinfo{title}{Estimation of local orientations in fibrous
  structures with applications to the purkinje system},
\newblock \bibinfo{journal}{IEEE Transactions on Biomedical Engineering}
  \bibinfo{volume}{58} (\bibinfo{year}{2011}) \bibinfo{pages}{1762--1772}.
  \DOIprefix\doi{10.1109/TBME.2011.2116119}.
\bibitem[{Costabal et~al.(2016)Costabal, Hurtado, and
  Kuhl}]{fractal_tree_citation}
\bibinfo{author}{F.~S. Costabal}, \bibinfo{author}{D.~E. Hurtado},
  \bibinfo{author}{E.~Kuhl},
\newblock \bibinfo{title}{Generating purkinje networks in the human heart},
\newblock \bibinfo{journal}{Journal of Biomechanics} \bibinfo{volume}{49}
  (\bibinfo{year}{2016}) \bibinfo{pages}{2455--2465}.
  \DOIprefix\doi{https://doi.org/10.1016/j.jbiomech.2015.12.025},
  \bibinfo{note}{cardiovascular Biomechanics in Health and Disease}.
\bibitem[{Ijiri et~al.(2008)Ijiri, Ashihara, Yamaguchi, Takayama, Igarashi,
  Shimada, Namba, Haraguchi, and Nakazawa}]{fractal_tree_methods}
\bibinfo{author}{T.~Ijiri}, \bibinfo{author}{T.~Ashihara},
  \bibinfo{author}{T.~Yamaguchi}, \bibinfo{author}{K.~Takayama},
  \bibinfo{author}{T.~Igarashi}, \bibinfo{author}{T.~Shimada},
  \bibinfo{author}{T.~Namba}, \bibinfo{author}{R.~Haraguchi},
  \bibinfo{author}{K.~Nakazawa},
\newblock \bibinfo{title}{A procedural method for modeling the purkinje fibers
  of the heart},
\newblock \bibinfo{journal}{The Journal of Physiological Sciences}
  \bibinfo{volume}{58} (\bibinfo{year}{2008}) \bibinfo{pages}{481--486}.
  \DOIprefix\doi{10.2170/physiolsci.RP003208}.
\bibitem[{Vergara et~al.(2014)Vergara, Palamara, Catanzariti, Nobile, Faggiano,
  Pangrazzi, Centonze, Maines, Quarteroni, and Vergara}]{using_fractal}
\bibinfo{author}{C.~Vergara}, \bibinfo{author}{S.~Palamara},
  \bibinfo{author}{D.~Catanzariti}, \bibinfo{author}{F.~Nobile},
  \bibinfo{author}{E.~Faggiano}, \bibinfo{author}{C.~Pangrazzi},
  \bibinfo{author}{M.~Centonze}, \bibinfo{author}{M.~Maines},
  \bibinfo{author}{A.~Quarteroni}, \bibinfo{author}{G.~Vergara},
\newblock \bibinfo{title}{Patient-specific generation of the purkinje network
  driven by clinical measurements of a normal propagation},
\newblock \bibinfo{journal}{Medical and Biological Engineering and Computing}
  \bibinfo{volume}{52} (\bibinfo{year}{2014}) \bibinfo{pages}{813--826}.
  \URLprefix \url{https://doi.org/10.1007/s11517-014-1183-5}.
  \DOIprefix\doi{10.1007/s11517-014-1183-5}.
\bibitem[{Zappon et~al.(2024)Zappon, Gsell, Gillette, and
  Plank}]{example_SA_ECG_OAT}
\bibinfo{author}{E.~Zappon}, \bibinfo{author}{M.~Gsell},
  \bibinfo{author}{K.~Gillette}, \bibinfo{author}{G.~Plank},
  \bibinfo{title}{Quantifying variabilities in cardiac digital twin models of
  the electrocardiogram}, \bibinfo{year}{2024}.
  \DOIprefix\doi{10.48550/arXiv.2407.17146}.
\bibitem[{Venton et~al.(2022)Venton, Gillette, Karli, Gsell, Loewe, Nagel,
  Winkler, and Wright}]{example_SA_ECG_OAT_2}
\bibinfo{author}{J.~Venton}, \bibinfo{author}{Gillette},
  \bibinfo{author}{Karli}, \bibinfo{author}{M.~Gsell},
  \bibinfo{author}{A.~Loewe}, \bibinfo{author}{C.~Nagel},
  \bibinfo{author}{B.~Winkler}, \bibinfo{author}{L.~Wright},
\newblock \bibinfo{title}{Sensitivity analysis of electrocardiogram features to
  computational model input parameters},
\newblock in: \bibinfo{booktitle}{2022 Computing in Cardiology (CinC)}, volume
  \bibinfo{volume}{498}, \bibinfo{year}{2022}, pp. \bibinfo{pages}{1--4}.
  \DOIprefix\doi{10.22489/CinC.2022.024}.
\bibitem[{S{\'a}nchez et~al.(2018)S{\'a}nchez, D'Ambrosio, Maffessanti, Caiani,
  Prinzen, Krause, Auricchio, and Potse}]{example_SA_ECG_OAT_3}
\bibinfo{author}{C.~S{\'a}nchez}, \bibinfo{author}{G.~D'Ambrosio},
  \bibinfo{author}{F.~Maffessanti}, \bibinfo{author}{E.~G. Caiani},
  \bibinfo{author}{F.~W. Prinzen}, \bibinfo{author}{R.~Krause},
  \bibinfo{author}{A.~Auricchio}, \bibinfo{author}{M.~Potse},
\newblock \bibinfo{title}{Sensitivity analysis of ventricular activation and
  electrocardiogram in tailored models of heart-failure patients},
\newblock \bibinfo{journal}{Medical \& Biological Engineering \& Computing}
  \bibinfo{volume}{56} (\bibinfo{year}{2018}) \bibinfo{pages}{491--504}.
  \DOIprefix\doi{10.1007/s11517-017-1696-9}, \bibinfo{note}{epub 2017 Aug 19}.
\bibitem[{Mincholé et~al.(2019)Mincholé, Zacur, Ariga, Grau, and
  Rodriguez}]{second_sigma_b}
\bibinfo{author}{A.~Mincholé}, \bibinfo{author}{E.~Zacur},
  \bibinfo{author}{R.~Ariga}, \bibinfo{author}{V.~Grau},
  \bibinfo{author}{B.~Rodriguez},
\newblock \bibinfo{title}{Mri-based computational torso/biventricular
  multiscale models to investigate the impact of anatomical variability on the
  ecg qrs complex},
\newblock \bibinfo{journal}{Frontiers in Physiology} \bibinfo{volume}{10}
  (\bibinfo{year}{2019}). \DOIprefix\doi{10.3389/fphys.2019.01103}.
\bibitem[{Cranford et~al.(2018)Cranford, O'Hara, Villongco, Hafez, Blake,
  Loscalzo, Fattebert, Richards, Zhang, Glosli, McCulloch, Krummen, Lightstone,
  and Wong}]{previous_HPS_SA}
\bibinfo{author}{J.~P. Cranford}, \bibinfo{author}{T.~J. O'Hara},
  \bibinfo{author}{C.~T. Villongco}, \bibinfo{author}{O.~M. Hafez},
  \bibinfo{author}{R.~C. Blake}, \bibinfo{author}{J.~Loscalzo},
  \bibinfo{author}{J.-L. Fattebert}, \bibinfo{author}{D.~F. Richards},
  \bibinfo{author}{X.~Zhang}, \bibinfo{author}{J.~N. Glosli},
  \bibinfo{author}{A.~D. McCulloch}, \bibinfo{author}{D.~E. Krummen},
  \bibinfo{author}{F.~C. Lightstone}, \bibinfo{author}{S.~E. Wong},
\newblock \bibinfo{title}{Efficient computational modeling of human ventricular
  activation and its electrocardiographic representation: A sensitivity study},
\newblock \bibinfo{journal}{Cardiovascular Engineering and Technology}
  \bibinfo{volume}{9} (\bibinfo{year}{2018}) \bibinfo{pages}{447--467}.
  \DOIprefix\doi{10.1007/s13239-018-0347-0}.
\bibitem[{Gillette et~al.(2021)Gillette, Gsell, Bouyssier, Prassl, Neic,
  Vigmond, and Plank}]{including_HPS_CDT}
\bibinfo{author}{K.~Gillette}, \bibinfo{author}{M.~A. Gsell},
  \bibinfo{author}{J.~Bouyssier}, \bibinfo{author}{A.~J. Prassl},
  \bibinfo{author}{A.~Neic}, \bibinfo{author}{E.~J. Vigmond},
  \bibinfo{author}{G.~Plank},
\newblock \bibinfo{title}{Automated framework for the inclusion of a
  his-purkinje system in cardiac digital twins of ventricular
  electrophysiology},
\newblock \bibinfo{journal}{Annals of Biomedical Engineering}
  \bibinfo{volume}{49} (\bibinfo{year}{2021}) \bibinfo{pages}{3143--3153}.
  \DOIprefix\doi{10.1007/s10439-021-02825-9}, \bibinfo{note}{epub 2021 Aug 24}.
\bibitem[{Pathmanathan et~al.(2019)Pathmanathan, Cordeiro, and
  Gray}]{Pras_GSA_1}
\bibinfo{author}{Pathmanathan}, \bibinfo{author}{Cordeiro},
  \bibinfo{author}{Gray},
\newblock \bibinfo{title}{Comprehensive uncertainty quantification and
  sensitivity analysis for cardiac action potential models},
\newblock \bibinfo{journal}{Frontiers in Physiology} \bibinfo{volume}{10}
  (\bibinfo{year}{2019}) \bibinfo{pages}{721}.
  \DOIprefix\doi{10.3389/fphys.2019.00721}.
\bibitem[{Davey et~al.(2024)Davey, Puelz, Rossi, Smith, Wells, Sturgeon,
  Segars, Vavalle, Peskin, and Griffith}]{davey_fsi_simulation}
\bibinfo{author}{M.~Davey}, \bibinfo{author}{C.~Puelz},
  \bibinfo{author}{S.~Rossi}, \bibinfo{author}{M.~A. Smith},
  \bibinfo{author}{D.~R. Wells}, \bibinfo{author}{G.~M. Sturgeon},
  \bibinfo{author}{W.~P. Segars}, \bibinfo{author}{J.~P. Vavalle},
  \bibinfo{author}{C.~S. Peskin}, \bibinfo{author}{B.~E. Griffith},
\newblock \bibinfo{title}{Simulating cardiac fluid dynamics in the human
  heart},
\newblock \bibinfo{journal}{PNAS Nexus} \bibinfo{volume}{3}
  (\bibinfo{year}{2024}) \bibinfo{pages}{pgae392}. \URLprefix
  \url{https://doi.org/10.1093/pnasnexus/pgae392}.
  \DOIprefix\doi{10.1093/pnasnexus/pgae392}.
\bibitem[{Azzouzi et~al.(2011)Azzouzi, Coudière, Turpault, and
  Zemzemi}]{Monodomain_model}
\bibinfo{author}{A.~Azzouzi}, \bibinfo{author}{Y.~Coudière},
  \bibinfo{author}{R.~Turpault}, \bibinfo{author}{N.~Zemzemi},
\newblock \bibinfo{title}{A mathematical model of the purkinje-muscle
  junctions},
\newblock \bibinfo{journal}{Mathematical Biosciences and Engineering}
  \bibinfo{volume}{8} (\bibinfo{year}{2011}) \bibinfo{pages}{915--930}.
  \DOIprefix\doi{10.3934/mbe.2011.8.915}.
\bibitem[{Abdala(2025)}]{LaryssaThesis}
\bibinfo{author}{L.~Abdala}, \bibinfo{title}{Electro-Fluid-Mechanical
  Computational Models of the Human Heart}, \bibinfo{type}{Ph.d. dissertation},
  The University of North Carolina at Chapel Hill, \bibinfo{address}{Chapel
  Hill, USA}, \bibinfo{year}{2025}.
\bibitem[{Nash and Panfilov(2004)}]{nash_panfilov_model}
\bibinfo{author}{M.~P. Nash}, \bibinfo{author}{A.~V. Panfilov},
\newblock \bibinfo{title}{Electromechanical model of excitable tissue to study
  reentrant cardiac arrhythmias},
\newblock \bibinfo{journal}{Progress in Biophysics and Molecular Biology}
  \bibinfo{volume}{85} (\bibinfo{year}{2004}) \bibinfo{pages}{501--522}.
  \DOIprefix\doi{https://doi.org/10.1016/j.pbiomolbio.2004.01.016},
  \bibinfo{note}{modelling Cellular and Tissue Function}.
\bibitem[{Izzo and Jackiewicz(2022)}]{ssprk2}
\bibinfo{author}{G.~Izzo}, \bibinfo{author}{Z.~Jackiewicz},
\newblock \bibinfo{title}{Strong stability preserving runge–kutta and linear
  multistep methods},
\newblock \bibinfo{journal}{Bulletin of the Iranian Mathematical Society}
  \bibinfo{volume}{48} (\bibinfo{year}{2022}) \bibinfo{pages}{4029--4062}.
  \URLprefix \url{https://doi.org/10.1007/s41980-022-00731-x}.
  \DOIprefix\doi{10.1007/s41980-022-00731-x}.
\bibitem[{Bishop and Plank(2011)}]{ecg_eq_1}
\bibinfo{author}{M.~J. Bishop}, \bibinfo{author}{G.~Plank},
\newblock \bibinfo{title}{Bidomain ecg simulations using an augmented
  monodomain model for the cardiac source},
\newblock \bibinfo{journal}{IEEE Transactions on Biomedical Engineering}
  \bibinfo{volume}{58} (\bibinfo{year}{2011}) \bibinfo{pages}{2297--2307}.
  \DOIprefix\doi{10.1109/TBME.2011.2148718}.
\bibitem[{Keller et~al.(2012)Keller, Schuler, Seemann, and Dössel}]{ecg_eq_2}
\bibinfo{author}{M.~W. Keller}, \bibinfo{author}{S.~Schuler},
  \bibinfo{author}{G.~Seemann}, \bibinfo{author}{O.~Dössel},
\newblock \bibinfo{title}{Differences in intracardiac signals on a realistic
  catheter geometry using mono- and bidomain models},
\newblock in: \bibinfo{booktitle}{2012 Computing in Cardiology},
  \bibinfo{year}{2012}, pp. \bibinfo{pages}{305--308}.
\bibitem[{Pilia et~al.(2021)Pilia, Nagel, Lenis, Becker, Dössel, and
  Loewe}]{ECGDeli}
\bibinfo{author}{N.~Pilia}, \bibinfo{author}{C.~Nagel},
  \bibinfo{author}{G.~Lenis}, \bibinfo{author}{S.~Becker},
  \bibinfo{author}{O.~Dössel}, \bibinfo{author}{A.~Loewe},
\newblock \bibinfo{title}{Ecgdeli - an open source ecg delineation toolbox for
  matlab},
\newblock \bibinfo{journal}{SoftwareX} \bibinfo{volume}{13}
  (\bibinfo{year}{2021}) \bibinfo{pages}{100639}.
  \DOIprefix\doi{https://doi.org/10.1016/j.softx.2020.100639}.
\bibitem[{Aguilar et~al.(2015)Aguilar, Albero, Albero, Rodriguez, Rodríguez,
  and Javier}]{torso_model}
\bibinfo{author}{S.~Aguilar}, \bibinfo{author}{R.~Albero},
  \bibinfo{author}{F.~Albero}, \bibinfo{author}{A.~Rodriguez},
  \bibinfo{author}{S.~Rodríguez}, \bibinfo{author}{F.~Javier},
  \bibinfo{title}{Human atria and torso 3d computational models for simulation
  of atrial arrhythmias}, \bibinfo{year}{2015}. \URLprefix
  \url{http://hdl.handle.net/10251/55150}.
  \DOIprefix\doi{10.4995/Dataset/10251/55150}.
\bibitem[{Ohlendorf et~al.(2020)Ohlendorf, Gerez, Porsch, Holzgreve, Maltry,
  Ackermann, and Groneberg}]{torso_dimensions}
\bibinfo{author}{D.~Ohlendorf}, \bibinfo{author}{A.~Gerez},
  \bibinfo{author}{L.~Porsch}, \bibinfo{author}{F.~Holzgreve},
  \bibinfo{author}{L.~Maltry}, \bibinfo{author}{H.~Ackermann},
  \bibinfo{author}{D.~A. Groneberg},
\newblock \bibinfo{title}{Standard reference values of the upper body posture
  in healthy male adults aged between 41 and 50 years in germany},
\newblock \bibinfo{journal}{Scientific Reports} \bibinfo{volume}{10}
  (\bibinfo{year}{2020}) \bibinfo{pages}{3823}.
  \DOIprefix\doi{10.1038/s41598-020-60813-w}.
\bibitem[{Khunti(2014)}]{ecg_lead_placement}
\bibinfo{author}{K.~Khunti},
\newblock \bibinfo{title}{Accurate interpretation of the 12-lead ecg electrode
  placement: A systematic review},
\newblock \bibinfo{journal}{Health Education Journal} \bibinfo{volume}{73}
  (\bibinfo{year}{2014}) \bibinfo{pages}{610--623}.
  \DOIprefix\doi{10.1177/0017896912472328}.
\bibitem[{Kligfield et~al.(2007)Kligfield, Gettes, Bailey, Childers, Deal,
  Hancock, van Herpen, Kors, Macfarlane, Mirvis, Pahlm, Rautaharju, and
  Wagner}]{aVL_formula}
\bibinfo{author}{P.~Kligfield}, \bibinfo{author}{L.~S. Gettes},
  \bibinfo{author}{J.~J. Bailey}, \bibinfo{author}{R.~Childers},
  \bibinfo{author}{B.~J. Deal}, \bibinfo{author}{E.~W. Hancock},
  \bibinfo{author}{G.~van Herpen}, \bibinfo{author}{J.~A. Kors},
  \bibinfo{author}{P.~Macfarlane}, \bibinfo{author}{D.~M. Mirvis},
  \bibinfo{author}{O.~Pahlm}, \bibinfo{author}{P.~Rautaharju},
  \bibinfo{author}{G.~S. Wagner},
\newblock \bibinfo{title}{Recommendations for the standardization and
  interpretation of the electrocardiogram},
\newblock \bibinfo{journal}{Circulation} \bibinfo{volume}{115}
  (\bibinfo{year}{2007}) \bibinfo{pages}{1306--1324}.
  \DOIprefix\doi{10.1161/CIRCULATIONAHA.106.180200}.
  \href{http://arxiv.org/abs/https://www.ahajournals.org/doi/pdf/10.1161/CIRCULATIONAHA.106.180200}{{\tt
  arXiv:https://www.ahajournals.org/doi/pdf/10.1161/CIRCULATIONAHA.106.180200}}.
\bibitem[{Tercero-Báez and Martín-Vaquero(2024)}]{IMEX}
\bibinfo{author}{A.~Tercero-Báez}, \bibinfo{author}{J.~Martín-Vaquero},
  \bibinfo{title}{On the stability of imex bdf methods for ddes and pddes},
  \bibinfo{year}{2024}. \URLprefix \url{https://arxiv.org/abs/2412.12297}.
  \href{http://arxiv.org/abs/2412.12297}{{\tt arXiv:2412.12297}}.
\bibitem[{Saltelli(2002)}]{saltelli_sampling}
\bibinfo{author}{A.~Saltelli},
\newblock \bibinfo{title}{Making best use of model evaluations to compute
  sensitivity indices},
\newblock \bibinfo{journal}{Computer Physics Communications}
  \bibinfo{volume}{145} (\bibinfo{year}{2002}) \bibinfo{pages}{280--297}.
  \DOIprefix\doi{10.1016/S0010-4655(02)00280-1}.
\bibitem[{Sobol'(1990)}]{sobol_indices_1}
\bibinfo{author}{I.~M. Sobol'},
\newblock \bibinfo{title}{On sensitivity estimation for nonlinear mathematical
  models},
\newblock \bibinfo{journal}{Mat. Model.} \bibinfo{volume}{2}
  (\bibinfo{year}{1990}) \bibinfo{pages}{112--118}.
\bibitem[{Saltelli et~al.(2008)Saltelli, Ratto, Andres, Campolongo, Cariboni,
  Gatelli, Saisana, and Tarantola}]{sobol_indices_2}
\bibinfo{author}{A.~Saltelli}, \bibinfo{author}{M.~Ratto},
  \bibinfo{author}{T.~Andres}, \bibinfo{author}{F.~Campolongo},
  \bibinfo{author}{J.~Cariboni}, \bibinfo{author}{D.~Gatelli},
  \bibinfo{author}{M.~Saisana}, \bibinfo{author}{S.~Tarantola},
  \bibinfo{title}{Global sensitivity analysis: The primer},
  \bibinfo{publisher}{John Wiley \& Sons, Ltd}, \bibinfo{address}{Chichester,
  UK}, \bibinfo{year}{2008}.
\bibitem[{Zhang et~al.(2015)Zhang, Trame, Lesko, and
  Schmidt}]{sobol_interpretation_guide}
\bibinfo{author}{X.~Y. Zhang}, \bibinfo{author}{M.~N. Trame},
  \bibinfo{author}{L.~J. Lesko}, \bibinfo{author}{S.~Schmidt},
\newblock \bibinfo{title}{Sobol sensitivity analysis: A tool to guide the
  development and evaluation of systems pharmacology models},
\newblock \bibinfo{journal}{CPT: Pharmacometrics \& Systems Pharmacology}
  \bibinfo{volume}{4} (\bibinfo{year}{2015}) \bibinfo{pages}{69--79}.
  \DOIprefix\doi{10.1002/psp4.6}.
\bibitem[{Iwanaga et~al.(2022)Iwanaga, Usher, and Herman}]{SALIB_1}
\bibinfo{author}{T.~Iwanaga}, \bibinfo{author}{W.~Usher},
  \bibinfo{author}{J.~Herman},
\newblock \bibinfo{title}{Toward {SALib} 2.0: {Advancing} the accessibility and
  interpretability of global sensitivity analyses},
\newblock \bibinfo{journal}{Socio-Environmental Systems Modelling}
  \bibinfo{volume}{4} (\bibinfo{year}{2022}) \bibinfo{pages}{18155}.
  \DOIprefix\doi{10.18174/sesmo.18155}.
\bibitem[{Herman and Usher(2017)}]{SALIB_2}
\bibinfo{author}{J.~Herman}, \bibinfo{author}{W.~Usher},
\newblock \bibinfo{title}{{SALib}: An open-source python library for
  sensitivity analysis},
\newblock \bibinfo{journal}{The Journal of Open Source Software}
  \bibinfo{volume}{2} (\bibinfo{year}{2017}).
  \DOIprefix\doi{10.21105/joss.00097}.
\bibitem[{Peguero et~al.(2017)Peguero, Presti, Perez, Issa, Brenes, and
  Tolentino}]{LVH_diagnosis_criteria}
\bibinfo{author}{J.~G. Peguero}, \bibinfo{author}{S.~L. Presti},
  \bibinfo{author}{J.~Perez}, \bibinfo{author}{O.~Issa}, \bibinfo{author}{J.~C.
  Brenes}, \bibinfo{author}{A.~Tolentino},
\newblock \bibinfo{title}{Electrocardiographic criteria for the diagnosis of
  left ventricular hypertrophy},
\newblock \bibinfo{journal}{Journal of the American College of Cardiology}
  \bibinfo{volume}{69} (\bibinfo{year}{2017}) \bibinfo{pages}{1694--1703}.
  \DOIprefix\doi{10.1016/j.jacc.2017.01.037}.
\bibitem[{Nesheiwat et~al.(2023)Nesheiwat, Goyal, and Jagtap}]{afib_diagnoses}
\bibinfo{author}{Z.~Nesheiwat}, \bibinfo{author}{A.~Goyal},
  \bibinfo{author}{M.~Jagtap}, \bibinfo{title}{Atrial Fibrillation}, StatPearls
  [Internet], \bibinfo{publisher}{StatPearls Publishing},
  \bibinfo{address}{Treasure Island, FL}, \bibinfo{year}{2023}. \URLprefix
  \url{https://www.ncbi.nlm.nih.gov/books/NBK526072/}.

\end{thebibliography}

\newpage
\beginsupplement
\setcounter{enumiv}{0}

\noindent \textbf{\LARGE SUPPLEMENTARY INFORMATION}
\section*{ECG Calculation Validation}
\label{sec:ecg_validation}


\begin{figure}[h!]
\centering
\includegraphics[width=0.9\textwidth]{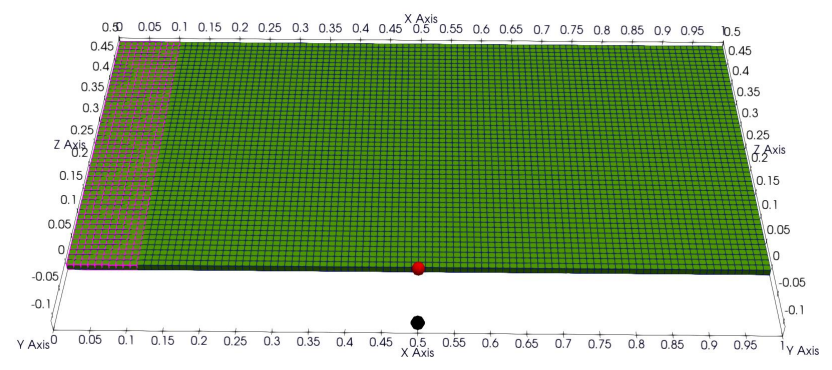} 
\caption{Tissue setup for the benchmark simulation. The red electrode corresponds to Electrode 2, while the black electrode corresponds to Electrode 1. A transmembrane current is applied to the pink region ($x\le0.1$ cm).}
\label{fig:benchmark_tissue} 
\end{figure}

\noindent To ensure the accuracy of ECG computations, we replicated a benchmark simulation presented in \citeS{ecg_eq_1} and compared our results. This validation step allowed us to confirm that the recovered ECG traces were computed correctly and exhibited the anticipated features. For this benchmark simulation, a myocardial tissue mesh with dimensions \(1.0 \times 0.01 \times 0.5 \, \text{cm}\) was generated using a regular hexahedral finite element grid, consisting of \(7,000\) elements and \(14,342\) nodes, see Figure \ref{fig:benchmark_tissue}. A transmembrane current pulse of \(50 \, \mu \text{A} / \text{cm}^2\) was applied over \(1 \, \mu \text{s}\) within the region \(x \leq 0.1 \, \text{cm}\), inducing propagation along the x-axis. The Ten Tusscher-Panfilov 2005 (TP06) ionic model was used to represent cell membrane dynamics for this benchmark study, with a resting membrane potential of \(-86.2 \, \text{mV}\) and a membrane capacitance of \(1 \, \mu \text{F} / \text{cm}^2\) \citeS{TP06_model}. Transmembrane voltages were computed using the monodomain model. Tissue conductivity values were defined as follows: bulk conductivity (\(\sigma_\text{b}\)) was set to \(10 \, \text{mS/cm}\), conductivity along the fiber direction (\(\sigma_\text{l}\)) was \(0.574 \, \text{mS/cm}\), and conductivity in the cross-fiber directions (\(\sigma_\text{n}\) and \(\sigma_\text{t}\)) was \(0.222 \, \text{mS/cm}\).  \\

\begin{figure}[h!]
\centering
\includegraphics[width=0.5\textwidth]{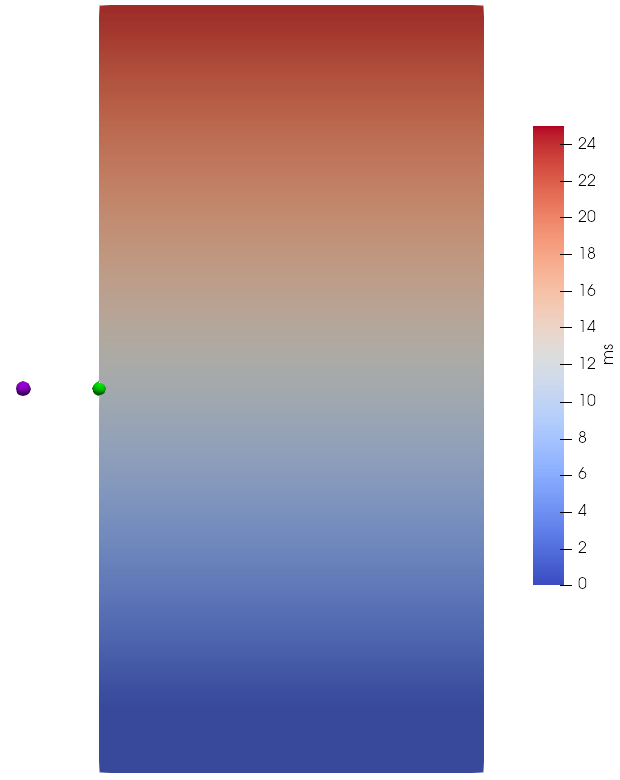} 
\caption{Activation map for benchmark simulation showing Electrode 1 (Purple) and Electrode 2 (Green). The entire mesh was activated within 25 ms, with the region around the electrodes activating at approximately 11 ms, corresponding to the peak observed in the ECG traces.}
\label{fig:activation_map_benchmark} 
\end{figure}

\begin{figure}[h!]
    \centering
    \begin{minipage}{0.5\textwidth}
        \centering
        \includegraphics[width=\linewidth]{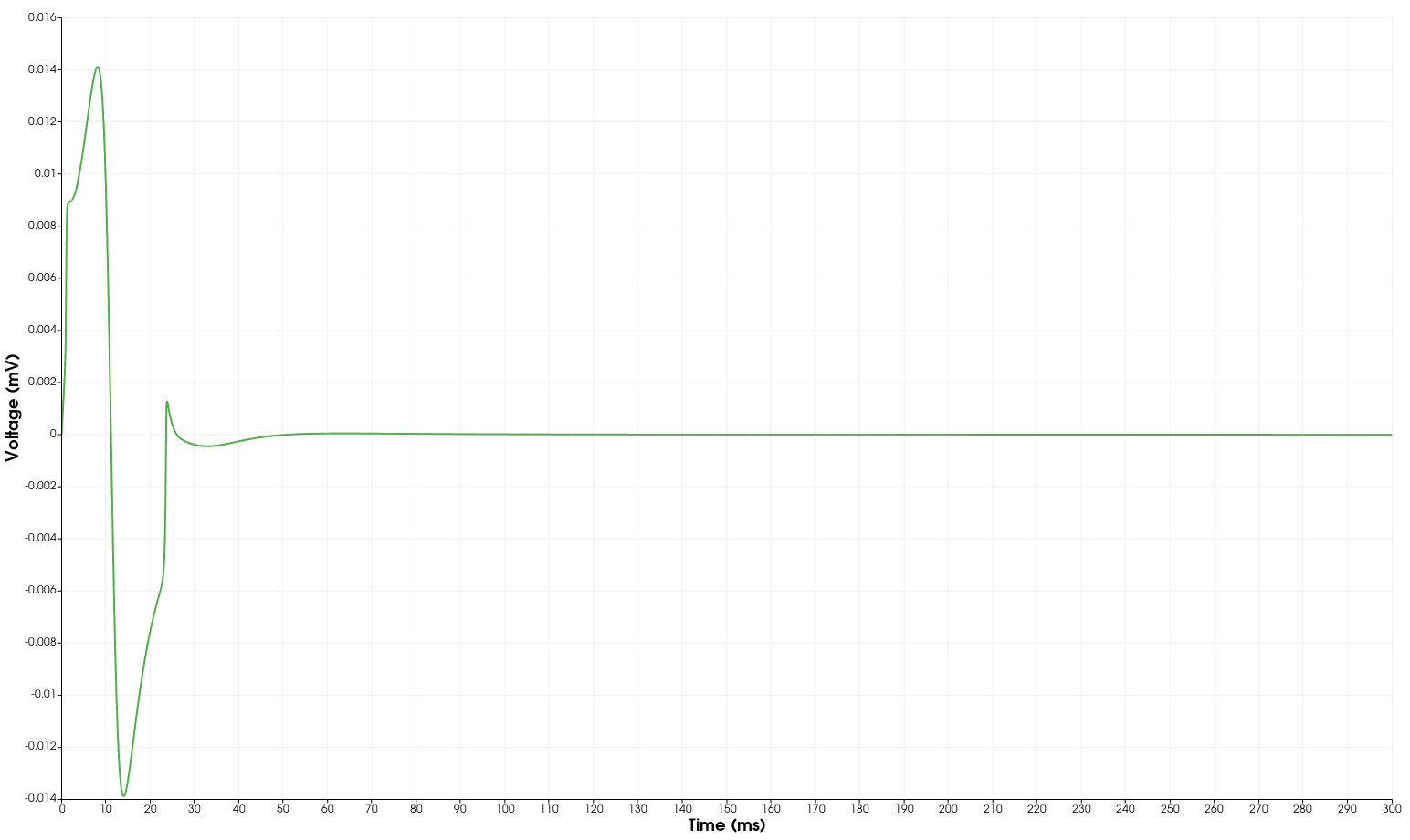}
    \end{minipage}%
    \begin{minipage}{0.5\textwidth}
        \centering
        \includegraphics[width=\linewidth]{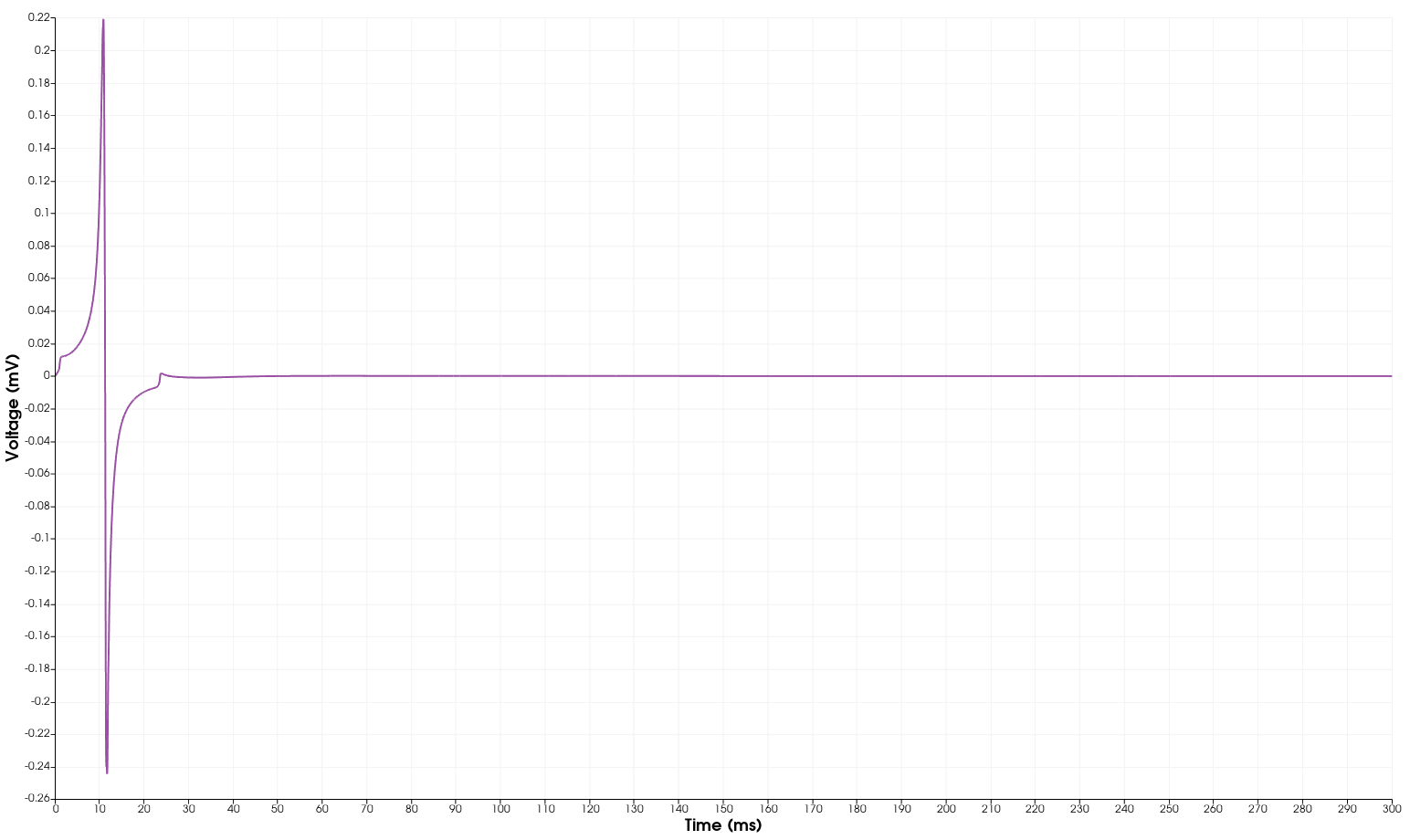}
    \end{minipage}%
    \caption{ECG Traces for Electrode 1 (Left) and 2 (Right) from benchmark simulation}
    \label{fig:benchmark_results}
\end{figure}

\noindent The first electrode was positioned \(0.1 \, \text{cm}\) away from the mesh in the y-direction, at the midpoint along both the x- and z-axes. The second electrode was placed at the same x- and z-coordinate but was directly on the tissue mesh. Figure \ref{fig:benchmark_results} displays the recovered ECG traces from this benchmark simulation. The simulation was run for a duration of 300 ms, with an ODE timestep of 0.03125 ms. The region around the electrode activated at approximately the same time as the peak seen in the ECG matching expected behavior (see Figure \ref{fig:activation_map_benchmark}). The overall pattern of the recovered traces aligns with the expected results and are consistent with similar previously conducted simulations \citeS{ecg_eq_1}.

\clearpage
\section*{Tables}

\begin{table}[h!]
    \centering
    \caption{Quantities of interest obtained from nominal simulation.}
    {\fontsize{11}{11}\selectfont 
    \renewcommand{\arraystretch}{2} 
    \setlength{\tabcolsep}{5pt} 
    \vspace{5 mm}
    \setlength{\tabcolsep}{50pt} 
    \begin{tabular}{@{}llc@{}}
        \toprule
        \textbf{Quantity of Interest}                 & \textbf{Value} \\
        \midrule
        Q-Wave Duration                    & 29.84 ms \\
        Q-Wave Peak-Time                   & 209.53 ms  \\
        Q-Wave Peak-Amplitude              & -0.56 mv     \\
        R-Wave Duration                    & 43.59 ms    \\
        R-Wave Peak-Time                   & 237.27 ms  \\
        R-Wave Peak-Amplitude              & 2.04 mv    \\
        S-Wave Duration                    & 29.61 ms    \\
        S-Wave Peak-Time                   & 276.09 ms  \\
        S-Wave Peak-Amplitude              & -0.62 mv   \\
        QRS Duration                       & 103.05 ms \\
        \bottomrule
    \end{tabular}
    \label{QOI_Nominal}
    }
\end{table}
\newpage

\subsection*{Standardized Quartiles for QOI}
\label{sec:standardized_quartiles}

\begin{table}[h!]
    \centering
    \caption{Standardized Quartiles $[z_{\text{Q1}}, z_{\text{Q3}}]$ for Peak Amplitudes across 4 Leads}
    {\fontsize{11}{11}\selectfont 
    \renewcommand{\arraystretch}{1.45} 
    \setlength{\tabcolsep}{3.5pt} 
    \begin{tabular}{lccc} 
        \toprule
        & \multicolumn{1}{c}{Q Wave} 
        & \multicolumn{1}{c}{R Wave} 
        & \multicolumn{1}{c}{S Wave} \\
        \midrule
        V2  & [-0.594, 0.500] & [-0.376, 0.522] & [-0.552, 0.556] \\
        V4  & [-0.584, 0.537] & [-0.610, 0.439] & [-0.526, 0.611] \\
        V6  & [-0.588, 0.568] & [-0.607, 0.440] & [-0.535, 0.612] \\
        aVL & [-0.627, 0.606] & [-0.413, 0.591] & [-0.614, 0.504] \\
        \bottomrule
    \end{tabular}
    }
    \label{tab:stats_amplitude_quartiles}
\end{table}

\begin{table}[h!]
    \centering
    \caption{Standardized Quartiles $[z_{\text{Q1}}, z_{\text{Q3}}]$ for Wave Durations across 4 Leads}
    {\fontsize{11}{11}\selectfont 
    \renewcommand{\arraystretch}{1.45} 
    \setlength{\tabcolsep}{3.5pt} 
    \begin{tabular}{lccc} 
        \toprule
        & \multicolumn{1}{c}{Q Wave} 
        & \multicolumn{1}{c}{R Wave} 
        & \multicolumn{1}{c}{S Wave} \\
        \midrule
        V2 & [-0.687,  0.560] & [-0.605, 0.523] &  [-0.408, 0.592] \\
        V4 & [-0.730, 0.563] & [-0.553, 0.511]  & [-0.450, 0.631]  \\
        V6 & [-0.716, 0.565] & [-0.579, 0.473] & [-0.444, 0.663]   \\
        aVL & [-0.661, 0.603] & [-0.400, 0.345] & [-0.356, 0.670]  \\
        \bottomrule
    \end{tabular}
    }
    \label{tab:stats_durations_quartiles}
\end{table}

\begin{table}[h!]
    \centering
    \caption{Standardized Quartiles $[z_{\text{Q1}}, z_{\text{Q3}}]$ for QRS Duration across 4 Leads}
    {\fontsize{11}{11}\selectfont 
    \renewcommand{\arraystretch}{1.45} 
    \setlength{\tabcolsep}{4.5pt} 
    \begin{tabular}{lc} 
        \toprule
        Lead & $[z_{Q1}, z_{Q3}]$ \\
        \midrule
        V2  & [-0.510, 0.637] \\
        V4  & [-0.572, 0.629] \\
        V6  & [-0.570, 0.657] \\
        aVL & [-0.554, 0.637] \\
        \bottomrule
    \end{tabular}
    }
    \label{tab:stats_QRS_duration_quartiles}
\end{table}

\begin{table}[h!]
    \centering
    \caption{Standardized Quartiles $[z_{\text{Q1}}, z_{\text{Q3}}]$ for Peak Times across 4 Leads}
    {\fontsize{11}{11}\selectfont 
    \renewcommand{\arraystretch}{1.45} 
    \setlength{\tabcolsep}{3.5pt} 
    \begin{tabular}{lccc} 
        \toprule
        & \multicolumn{1}{c}{Q Wave} 
        & \multicolumn{1}{c}{R Wave} 
        & \multicolumn{1}{c}{S Wave} \\
        \midrule
        V2 & [-0.476, 0.645] & [-0.434, 0.607] & [-0.468, 0.579] \\
        V4 & [-0.549, 0.625] & [-0.590, 0.602] & [-0.460, 0.615] \\
        V6 & [-0.562, 0.628] & [-0.605, 0.600] & [-0.463, 0.631] \\
        aVL & [-0.456, 0.726] & [-0.433, 0.629] & [-0.459, 0.688] \\
        \bottomrule
    \end{tabular}
    }
    \label{tab:stats_times_quartiles}
\end{table}

\clearpage
\subsection*{Significant Sobol Indices}

\begin{table}[h!]
    \centering
    \caption{Largest First-Order Sobol Indices for Wave Durations}
    {\fontsize{11}{11}\selectfont 
    \renewcommand{\arraystretch}{1.45} 
    \setlength{\tabcolsep}{5pt} 
    \begin{tabular}{l c c c c} 
        \toprule
        Parameter (QOI) & V2 & V4 & V6 & aVL \\
        \midrule
        First Fascicle Angle (QRS duration)  & 0.0389& 0.0363  & 0.0395 & 0.0528\\
        Median Branch Length (R duration)  & 0.0503& 0.0205& 0.0183& 0.0675 \\
        Branch Angle (R duration)  & 0.0349  & 0.00848  & 0.00871 & 0.0747\\
        \bottomrule
    \end{tabular}
    }
    \label{tab:first_order_wave_durations}
\end{table}

\begin{table}[h!]
    \centering
    \caption{Largest First-Order Sobol Indices for Peak Times}
    {\fontsize{11}{11}\selectfont 
    \renewcommand{\arraystretch}{1.45} 
    \setlength{\tabcolsep}{5pt} 
    \begin{tabular}{l c c c c} 
        \toprule
        Parameter (QOI) & V2 & V4 & V6 & aVL \\
        \midrule
        Number of Branches (Q Peak Time)  & 0.0910& 0.0913& 0.0913& 0.0935\\
        Number of Branches (R Peak Time)  & 0.0891& 0.0853& 0.0843& 0.101\\
        Number of Branches (S Peak Time)  & 0.0429& 0.0519& 0.0544& 0.0835\\
        \bottomrule
    \end{tabular}
    }
    \label{tab:first_order_sobol_times}
\end{table}

\begin{table}[h!]
    \centering
    \caption{Largest First-Order Sobol Indices for Peak Amplitudes}
    {\fontsize{11}{11}\selectfont 
    \renewcommand{\arraystretch}{1.45} 
    \setlength{\tabcolsep}{5pt} 
    \begin{tabular}{l c c c c} 
        \toprule
        Parameter (QOI) & V2 & V4 & V6 & aVL \\
        \midrule
        First Fascicle Angle (Q Peak Amplitude)  & 0.0205  & 0.0280& 0.0297 & 0.0406 \\
        First Fascicle Length (Q Peak Amplitude)  & 0.0271& 0.0277& 0.0272& 0.0465 \\
        First Fascicle Length (R Peak Amplitude)  & 0.0341& 0.0206& 0.0208& 0.0472\\
        Second Fascicle Length (R Peak Amplitude)  & 0.0458& 0.0436& 0.0425& 0.0474\\
        \bottomrule
    \end{tabular}
    }
    \label{tab:first_order_sobol_ampl}
\end{table}

\begin{table}[h!]
    \centering
    \caption{Total Sobol Sensitivity Indices for Peak Times}
    {\fontsize{9}{11}\selectfont 
    \renewcommand{\arraystretch}{2} 
    \setlength{\tabcolsep}{4pt} 
    \begin{tabular}{l *{12}{c}} 
        \toprule
        & \multicolumn{4}{c}{Q Wave} 
        & \multicolumn{4}{c}{R Wave} 
        & \multicolumn{4}{c}{S Wave} \\
        \cmidrule(lr){2-5} \cmidrule(lr){6-9} \cmidrule(lr){10-13}
        & V2 & V4 & V6 & aVL
        & V2 & V4 & V6 & aVL
        & V2 & V4 & V6 & aVL \\
        \midrule
        Number of Branches & 0.422 & 0.436 & 0.441 & 0.432
        & 0.408 & 0.465 & 0.469 & 0.422  
            & 0.415 & 0.429 & 0.431 & 0.428 \\
        Median Branch Length & 0.560 & 0.562 & 0.564 & 0.579 
            & 0.549 & 0.576 & 0.581 & 0.555 
            & 0.613 & 0.613 & 0.611 & 0.578 \\
        Branch Angle & 0.870 & 0.865 & 0.864 & 0.882  
            & 0.852 & 0.860 & 0.870 & 0.848  
            & 0.914 & 0.906 & 0.910 & 0.862 \\
        Repulsivity & 0.707 & 0.707 & 0.708 & 0.718 
            & 0.734 & 0.735 & 0.732 & 0.723 
            & 0.760 & 0.758 & 0.752 & 0.727\\
        Length of Segments in Branch & 0.351 & 0.362 & 0.367 & 0.365 
            & 0.343 & 0.398 & 0.401 & 0.348 
            & 0.386 & 0.391 & 0.390 & 0.365 \\
        First Fascicle Angle & 0.464 & 0.473 & 0.477 & 0.475 
            & 0.472 & 0.511 & 0.512 & 0.467 
            & 0.513 & 0.518 & 0.515 & 0.481\\
        Second Fascicle Angle & 0.868 & 0.866 & 0.866 & 0.862 
            & 0.855 & 0.872 & 0.876 & 0.843 
            & 0.925 & 0.917 & 0.914 & 0.863 \\
        First Fascicle Length & 0.597 & 0.600 & 0.602 & 0.608
            & 0.582 & 0.616 & 0.621 & 0.578 
            & 0.661 & 0.657 & 0.657 & 0.612 \\
        Second Fascicle Length & 0.487 & 0.499 & 0.503 & 0.488 
            & 0.490 & 0.534 & 0.535 & 0.485 
            & 0.498 & 0.509 & 0.507 & 0.493 \\
        \bottomrule
    \end{tabular}
    }
    \label{tab:total_sobol_indices_times}
\end{table}

\clearpage
\section*{Figures}
\subsection*{QOI Distributions}
\label{sec:QOI_distributions}
\begin{figure}[h!]
    \centering
    \begin{tabular}{cc}
        \includegraphics[width=0.5\textwidth]{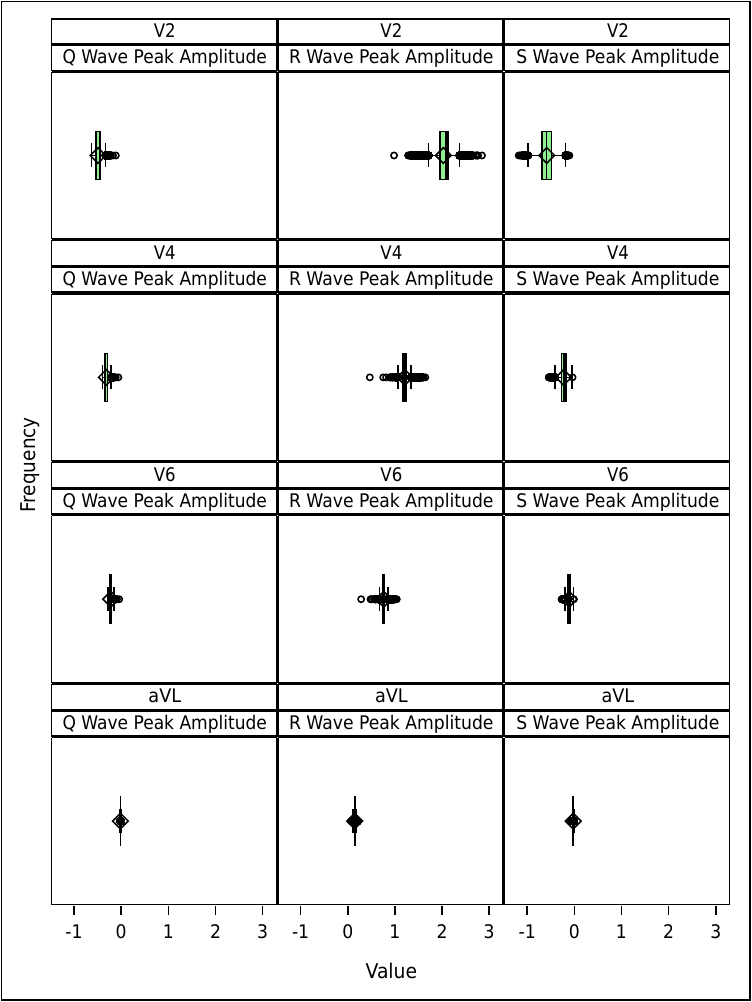}  & \includegraphics[width=0.5\textwidth]{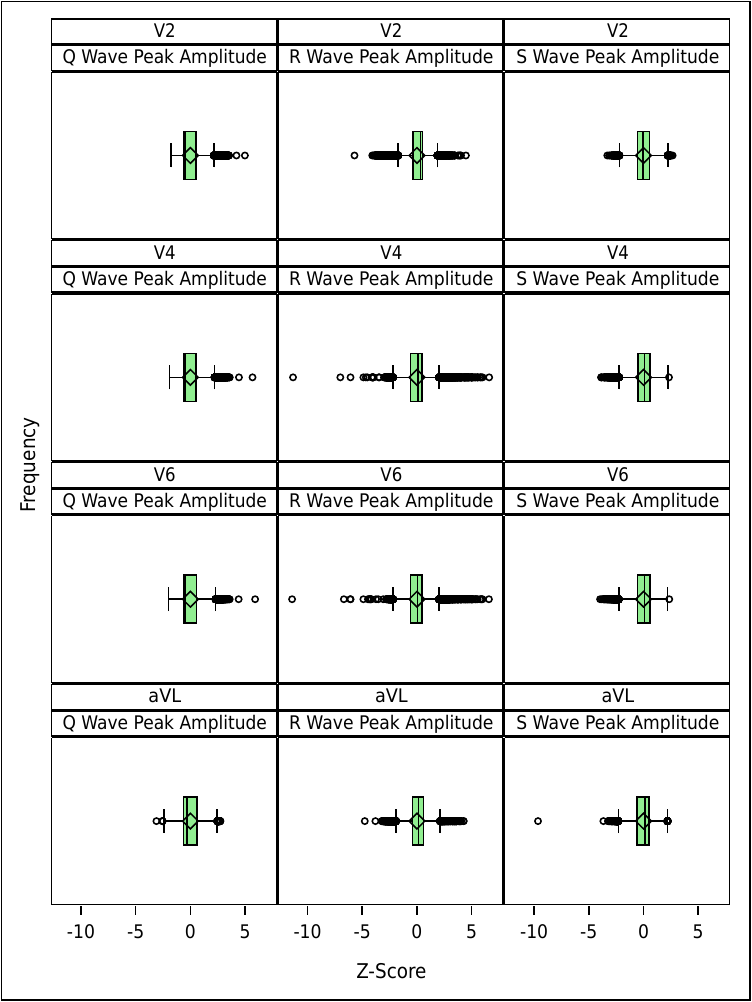} \\
    \end{tabular}
    \caption{Original (left) and z-score (right) distributions of Q, R, and S peak amplitudes (mV) across four electrode leads, based on 22,528 trials. Whiskers represent the minimum and maximum values within 1.5×IQR.}
    \label{fig:amplitude_distributions} 
\end{figure}
\newpage
\begin{figure}[ht]
    \centering
    \begin{tabular}{cc}
        \includegraphics[width=0.5\textwidth]{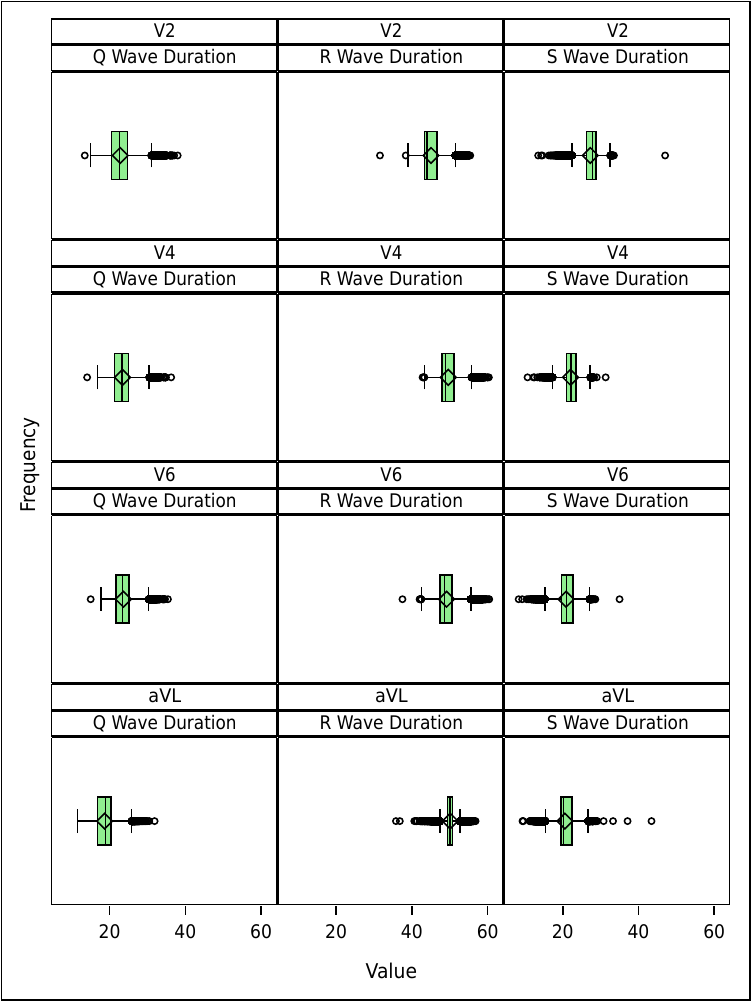}  & \includegraphics[width=0.5\textwidth]{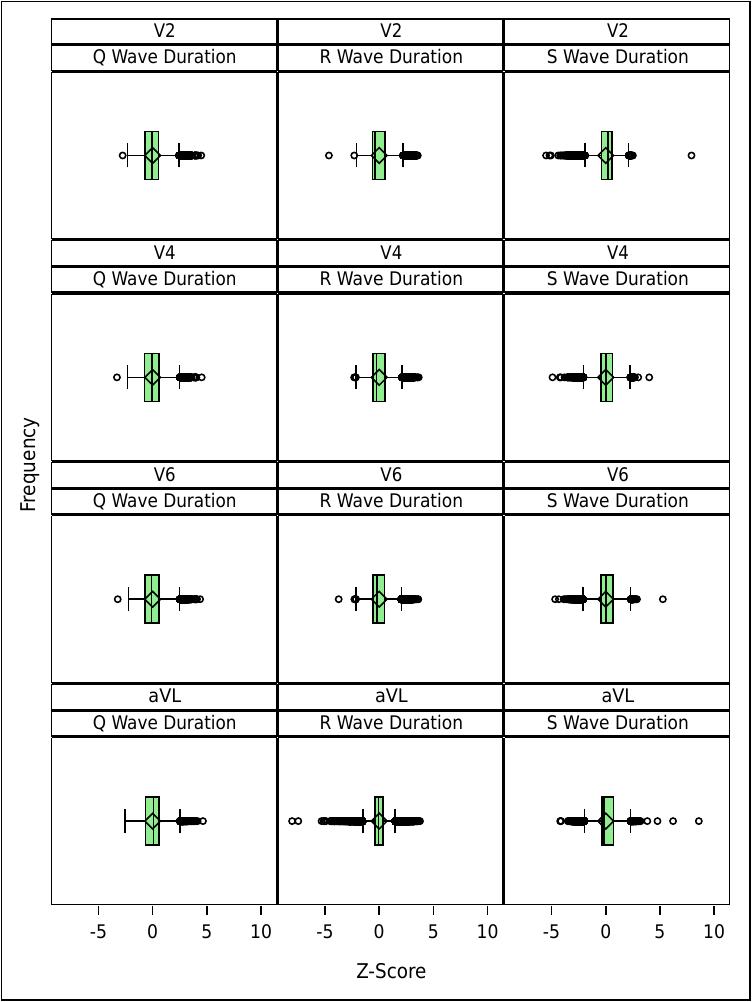} \\
    \end{tabular}
    \caption{Original (left) and z-score (right) distributions of Q, R, and S wave durations (ms) across four electrode leads, based on 22,528 trials. Whiskers represent the minimum and maximum values within 1.5×IQR.}
    \label{fig:wave_duration_distributions}
\end{figure}
\newpage
\begin{figure}[ht]
    \centering
    \begin{tabular}{cc}
        \includegraphics[width=0.5\textwidth]{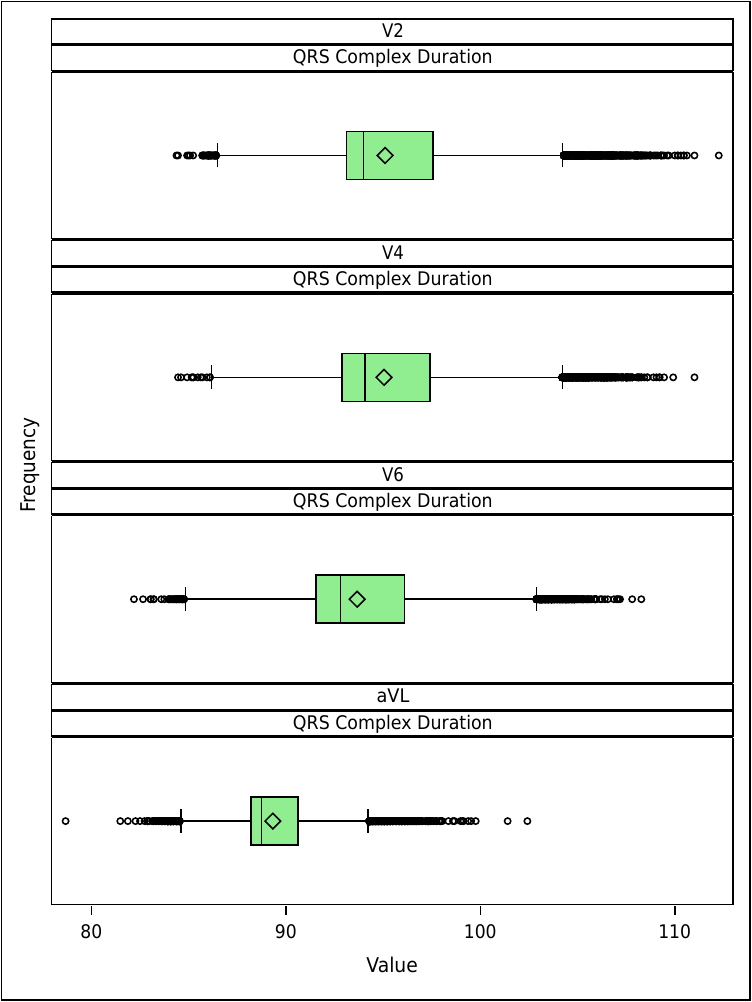}  & \includegraphics[width=0.5\textwidth]{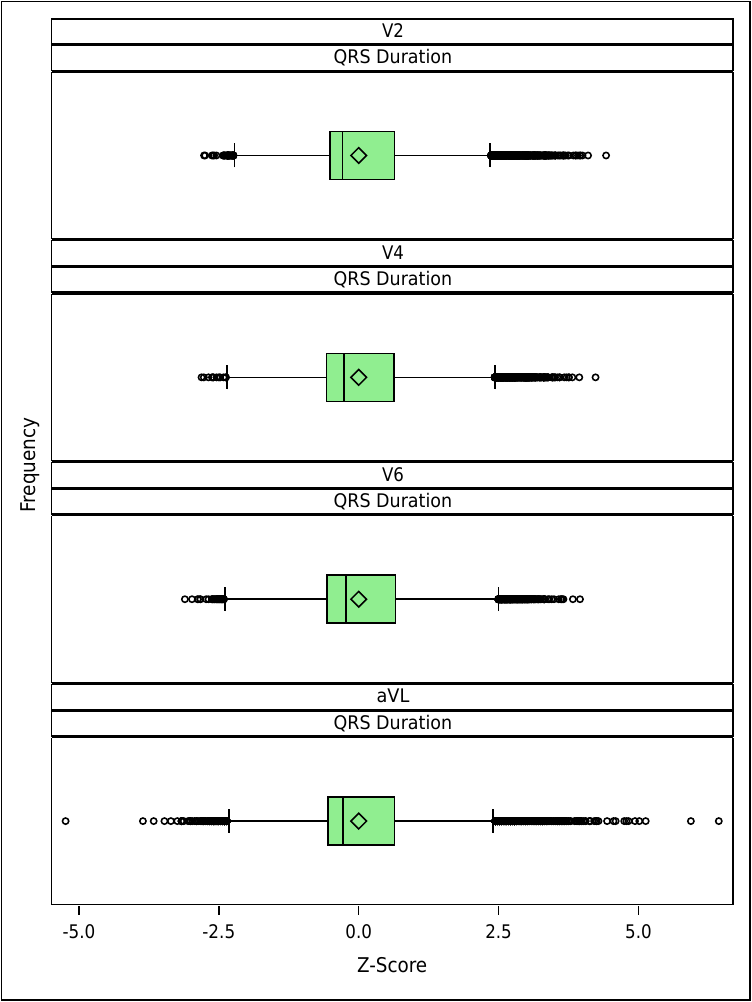} \\
    \end{tabular}
    \caption{Original (left) and z-score (right) distributions of QRS waveform durations (ms) across four electrode leads, based on 22,528 trials. Whiskers represent the minimum and maximum values within 1.5×IQR.}
    \label{fig:QRS_duration_distribution} 
\end{figure}
\newpage
\begin{figure}[h!]
    \centering
    \begin{tabular}{cc}
        \includegraphics[width=0.5\textwidth]{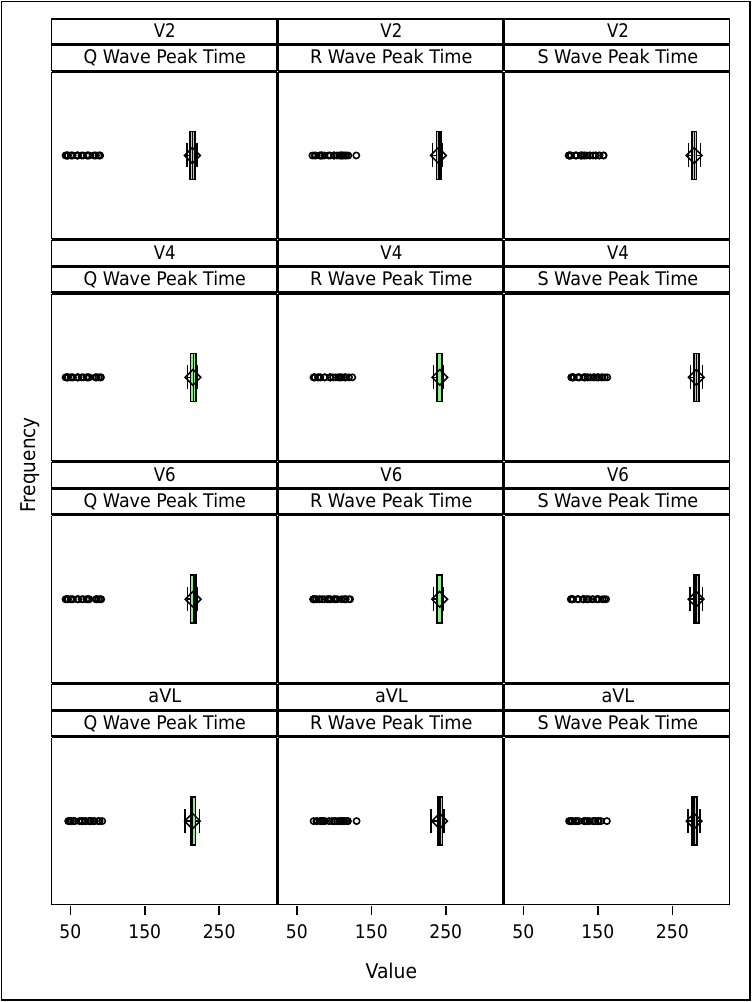}  & \includegraphics[width=0.5\textwidth]{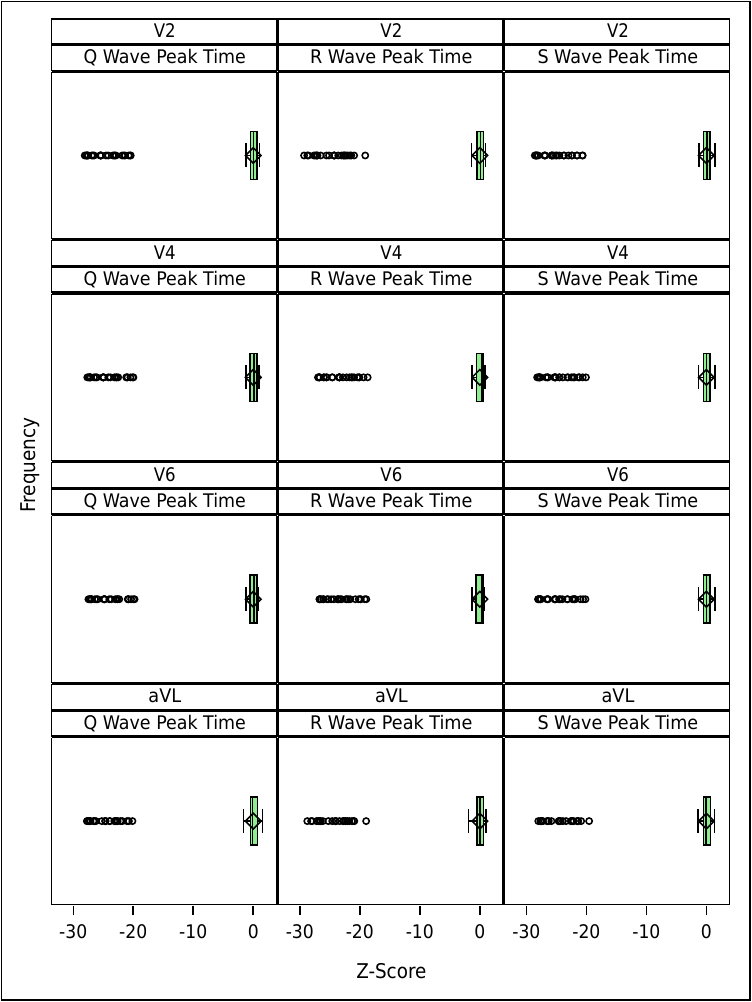} \\
    \end{tabular}
    \caption{Original (left) and z-score (right) distributions of Q, R, and S peak times (ms) across four electrode leads, based on 22,528 trials. Whiskers represent the minimum and maximum values within 1.5×IQR.}
    \label{fig:wave_time_distribution} 
\end{figure}
\newpage
\subsection*{Sobol Sensitivity Indices}
\newpage

\newpage
\subsection*{Outlier Trials}
\begin{figure}[h!]
    \centering
    \begin{tabular}{cc}
        \includegraphics[width=0.5\textwidth]{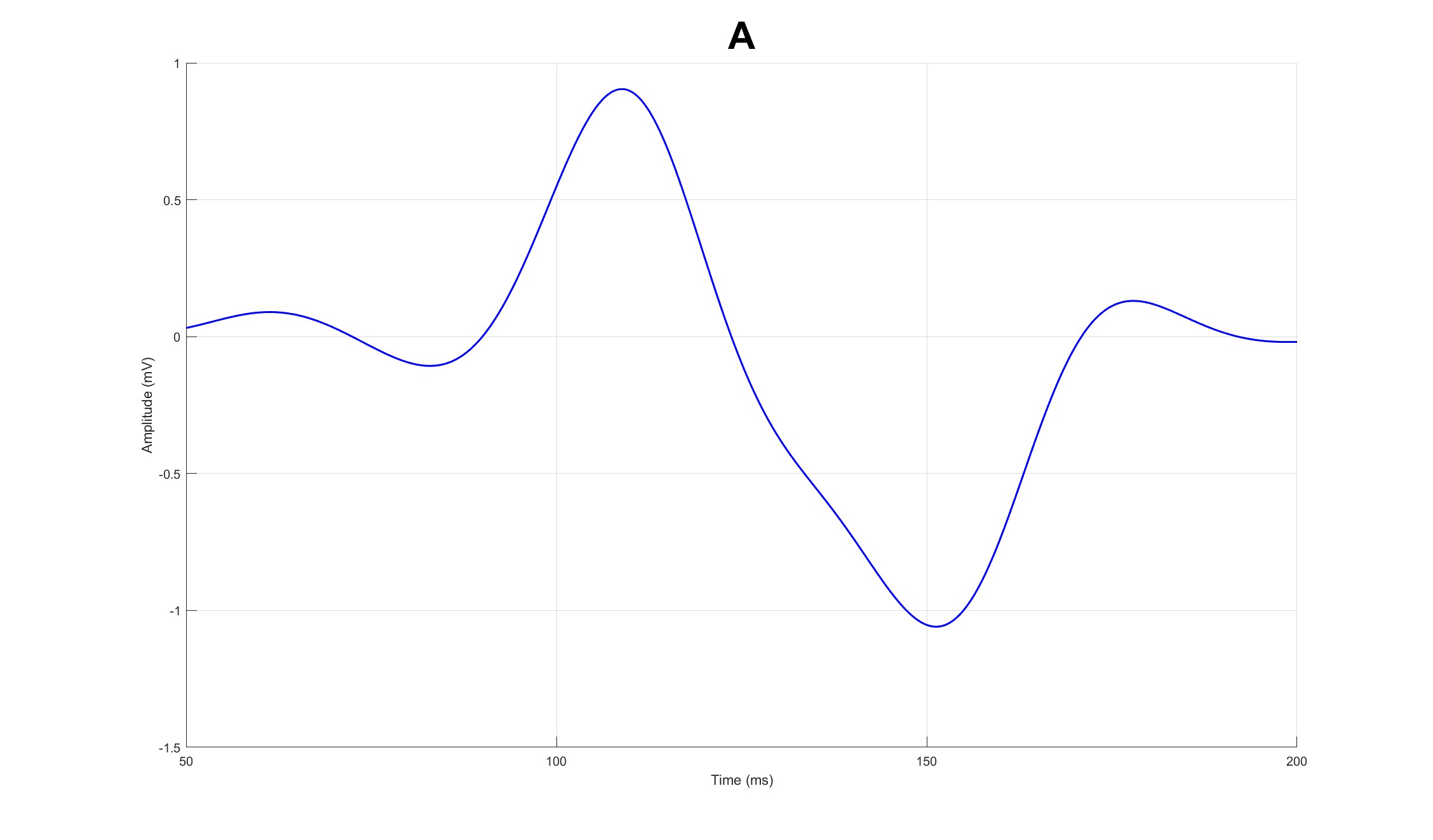} & \includegraphics[width=0.5\textwidth]{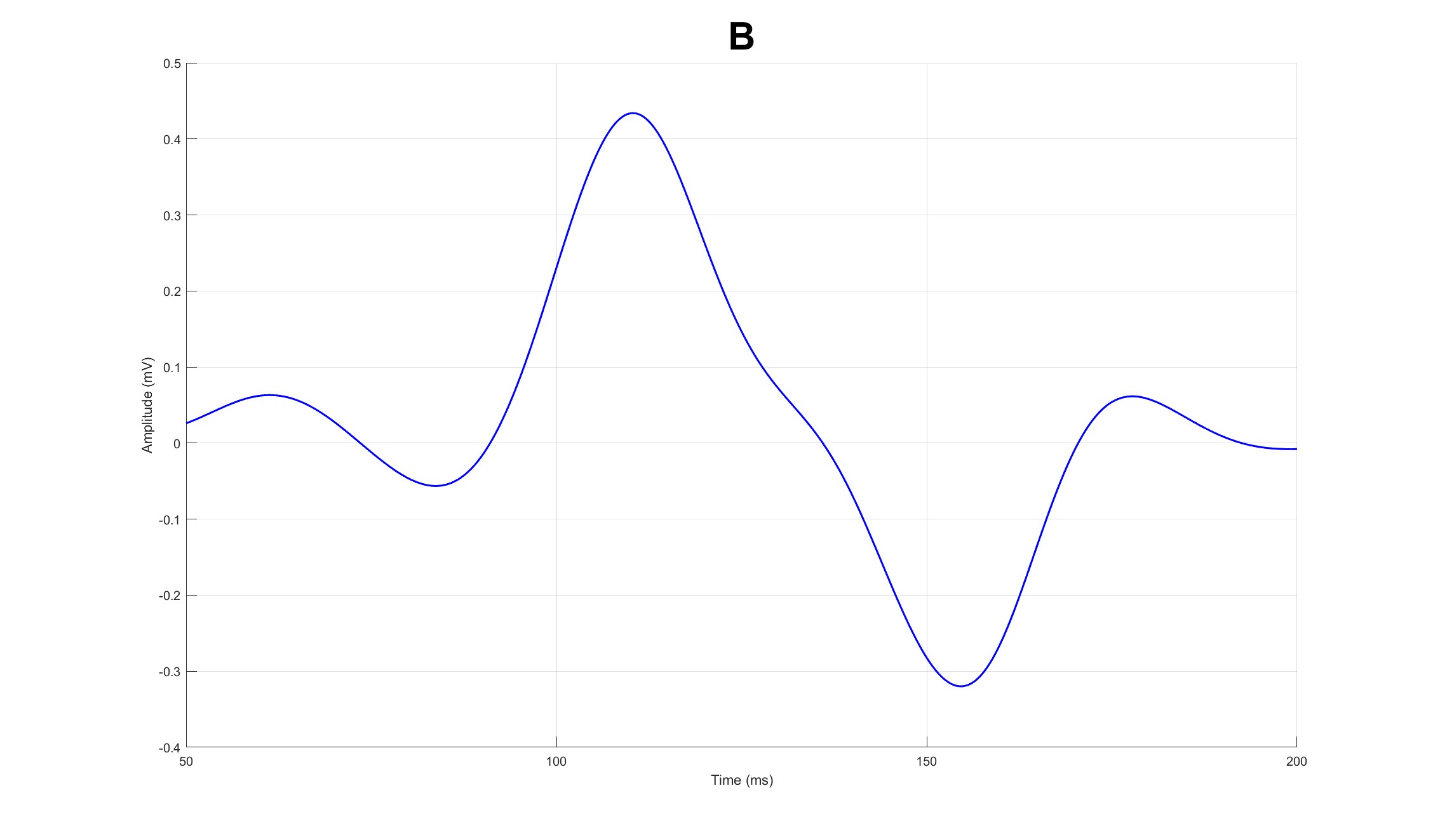} \\
        \includegraphics[width=0.5\textwidth]{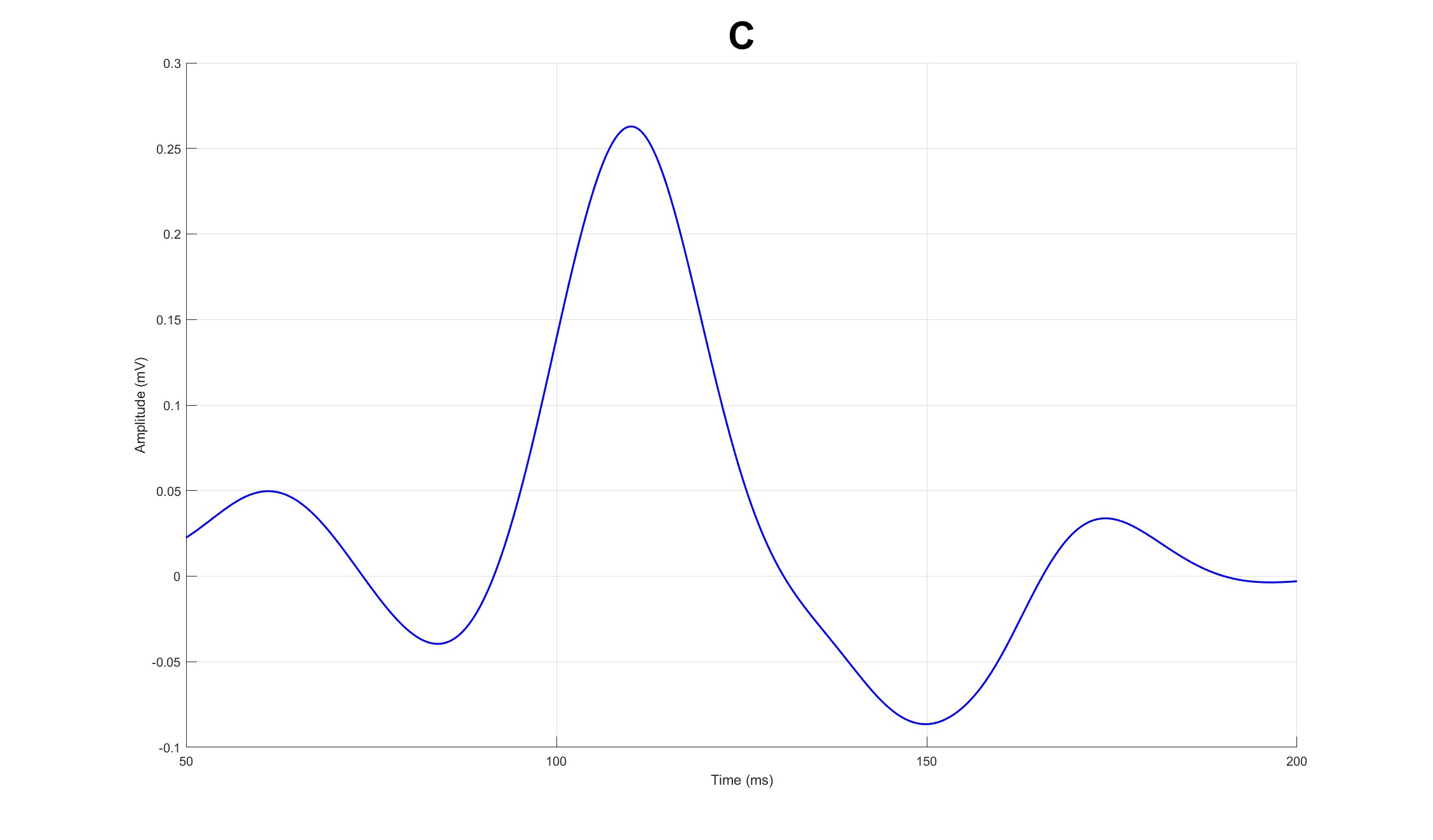} & \includegraphics[width=0.5\textwidth]{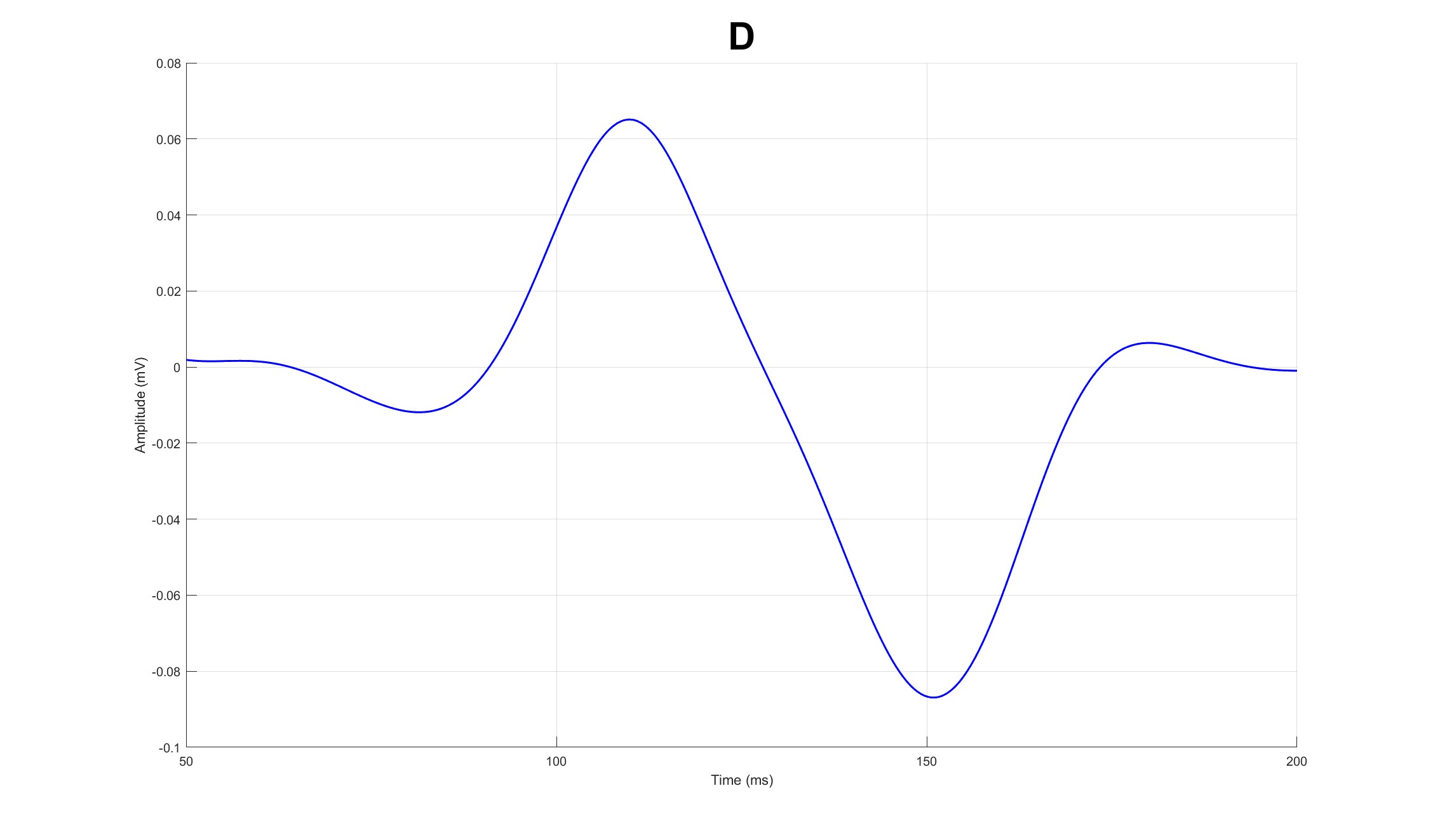} \\
    \end{tabular}
    \caption{QRS complex in leads V2 (A), V4 (B), V6 (C), and aVL (D) for outlier trial. The corresponding Purkinje network parameters are as follows: number of branches = 12, branch length = 5.96 mm, branch angle = 0.171 rad, repulsivity = 0.099, length of segments in branch = 0.121 mm, fascicle angle 1 = -0.092 rad, fascicle angle 2 = 0.251 rad, fascicle length 1 = 3.75 mm, and fascicle length 2 = 6.24 mm.}
    \label{fig:outlier_trial}
\end{figure}
\newpage

\clearpage
\renewcommand\refname{SUPPLEMENTAL REFERENCES}
\bibliographystyleS{elsarticle-num-names}
\bibliographyS{main}

\end{document}